\documentclass[a4paper,fleqn]{cas-dc}

\usepackage[authoryear]{natbib}

\usepackage{booktabs}
\usepackage{amssymb}
\usepackage{amsmath}
\usepackage{pifont}
\usepackage{multirow}
\usepackage{soul}
\usepackage[capitalize]{cleveref}

\def\tsc#1{\csdef{#1}{\textsc{\lowercase{#1}}\xspace}}
\tsc{WGM}
\tsc{QE}
\tsc{EP}
\tsc{PMS}
\tsc{BEC}
\tsc{DE}

\begin{document}
\let\WriteBookmarks\relax
\def\floatpagepagefraction{1}
\def\textpagefraction{.001}

\shorttitle{}

\shortauthors{T. Stegm\"uller et~al.}


\title [mode = title]{Self-supervised learning-based cervical cytology for the triage of HPV-positive women in resource-limited settings and low-data regime}

\let\printorcid\relax 
\author[1]{Thomas Stegm\"uller}
\cormark[1]
\author[1]{ Christian Abbet}
\author[1,3]{ Behzad Bozorgtabar}
\author[2]{ Holly Clarke}
\author[2]{ Patrick Petignat}
\author[2]{ Pierre Vassilakos}
\author[1,3]{ Jean-Philippe Thiran}

\affiliation[1]{organization={Ecole Polytechnique Fédérale de Lausanne},
    city={Lausanne},
    postcode={1015}, 
    country={Switzerland}}
    
\affiliation[2]{organization={Hôpitaux Universitaires de Genève},
    city={Genève},
    postcode={1205}, 
    country={Switzerland}}

\affiliation[3]{organization={Centre Hospitalier Universitaire Vaudois},
    city={Lausanne},
    postcode={1011},
    country={Switzerland}}

\cortext[cor1]{Corresponding author at: \texttt{thomas.stegmuller@epfl.ch}}

\crefname{section}{Sec.}{Secs.}
\Crefname{section}{Section}{Sections}
\Crefname{table}{Table}{Tables}
\crefname{table}{Tab.}{Tabs.}

\newcommand{\Thomas}[1]{\textcolor{blue}{#1}}
\newcommand{\Behzad}[1]{\textcolor{red}{#1}}

\newcommand{\aug}{\tilde{\boldsymbol{x}}}
\newcommand{\im}{\boldsymbol{x}}
\newcommand{\glob}{\bar{\boldsymbol{z}}}
\newcommand{\cls}{\texttt{[CLS]}}
\newcommand{\model}{g}
\newcommand{\backbone}{f}
\newcommand{\head}{h}
\newcommand{\bag}{\boldsymbol{X}}
\newcommand{\instance}{\boldsymbol{x}}
\newcommand{\ilabel}{y}
\newcommand{\blabel}{Y}

\newcommand{\vit}{\text{ViT}}
\newcommand{\resnet}{\text{ResNet}}
\newcommand{\abmil}{\text{AbMIL}}
\newcommand{\transmil}{\text{TransMIL}}
\newcommand{\clam}{\text{CLAM}}
\newcommand{\dino}{\text{DINO}}
\newcommand{\imagenet}{\text{ImageNet-1k}}
\newcommand{\overbar}[1]{\mkern 1.5mu\overline{\mkern-1.5mu#1\mkern-1.5mu}\mkern 1.5mu}
\newcommand{\supp}{\textbf{Supplementary Material}}

\newcommand{\pname}{\texttt{C}^{3}\texttt{P}}

\newcommand{\ck}{\textcolor{green!80!black}{\ding{51}}}
\newcommand{\xk}{\textcolor{red}{\ding{55}}}
\newcommand*\rot{\rotatebox{90}}

\definecolor{light_cyan}{HTML}{c6effc}
\sethlcolor{light_cyan}
\definecolor{dg}{rgb}{0, 0.5, 0}
\newcolumntype{g}{>{\columncolor{light_gray}}c}

\newcommand{\ie}{\textit{i}.\textit{e}., }
\newcommand{\eg}{\textit{e}.\textit{g}., }

\begin{abstract}
Screening Papanicolaou test samples has proven to be highly effective in reducing cervical cancer-related mortality. However, the lack of trained cytopathologists hinders its widespread implementation in low-resource settings. Deep learning-based telecytology diagnosis emerges as an appealing alternative, but it requires the collection of large annotated training datasets, which is costly and time-consuming. In this paper, we demonstrate that the abundance of unlabeled images that can be extracted from Pap smear test whole slide images presents a fertile ground for self-supervised learning methods, yielding performance improvements relative to readily available pre-trained models for various downstream tasks. In particular, we propose \textbf{C}ervical \textbf{C}ell \textbf{C}opy-\textbf{P}asting ($\pname$) as an effective augmentation method, which enables knowledge transfer from open-source and labeled single-cell datasets to unlabeled tiles. Not only does $\pname$ outperforms naive transfer from single-cell images, but we also demonstrate its advantageous integration into multiple instance learning methods. Importantly, all our experiments are conducted on our introduced  \textit{in-house} dataset comprising liquid-based cytology Pap smear images obtained using low-cost technologies. This aligns with our objective of leveraging deep learning-based telecytology for diagnosis in low-resource settings.
\end{abstract}

\begin{keywords}
Digital Cytology \sep WSIs Classification \sep Copy-Paste.
\end{keywords}

\maketitle
\section{Introduction}
\label{sec:introduction_new}
Cervical cancer is considered nearly completely preventable but continues to be a leading cause of cancer mortality. In 2020, about 342 000 women died from this disease, most of them in developing countries where cytology-based screening programs to detect and treat precancerous lesions are not available or affordable \cite{sung2021global}.\par
In the knowledge that HPV is the etiological factor that drives cervical cancer development, secondary prevention with human papillomavirus (HPV) testing has, in recent years, become the preferred screening method in many high-income settings. It is recommended by the WHO for women aged >30 years in low-and-middle-income countries (LMICs) \cite{world2021guideline}. Its high sensitivity and negative predictive value in detecting cervical intraepithelial neoplasia grade 2 or worse ($\geq$CIN2) allow extended screening intervals. Recently, the development of fully automated diagnostic devices providing rapid HPV testing of self-obtained vaginal samples has offered a great opportunity to improve the effectiveness of cervical cancer prevention in low-resource contexts \cite{saidu2021performance}.
 \par
However, a single HPV test has limited specificity and can lead to unnecessary workup and overtreatment. Therefore, a triage strategy is required for HPV-positive women to mitigate this difficulty. Cytology is generally proposed as it is an effective method for triaging HPV-positive women \cite{von2015european}, but in low-resource settings, various logistic and operational reasons prevent successful cytology implementation. Amongst other barriers, cytological triage can be time-consuming, and in countries that use cytology as a triage method, results are typically unavailable on the same day as sample collection. In lower-income settings, loss to follow-up means that this becomes a seriously limiting problem. In these settings, therefore, rapid tests that give same-day results and lead to decisions about treatment are preferred. 
\par
A solution for countries with limited resources could be affordable digital imaging technology for real-time remote cytologic diagnosis by specialists \cite{vassilakos2023telecytologic}. Using this scheme, the preparation and digitization of cervical smears from HPV-positive women would be performed on-site during the same visit using a ``test-triage-and-treat'' approach (3T-approach) \cite{levy2020implementing}. This process eliminates the need for in-house cytopathologists and might allow for reliable, cost-effective triage of HPV-positive women. Furthermore, to facilitate the visual analysis of Pap slides and reduce the screening time, deep learning-based algorithms could be used to obtain a rapid and accurate cytological diagnosis allowing a ``same-day treatment''. \par

The emergence of affordable and portable high-resolution scanners, such as the Grundium Ocus\textsuperscript{\textregistered}40, along with low-cost slide preparation procedures like SurePath\textsuperscript{\texttrademark}, creates a favorable environment for this endeavor. When it comes to the learning algorithm, the main expenses are associated with the annotation process and the level of expertise it demands. Nevertheless, acquiring a large and well-curated annotated dataset proves challenging and time-consuming. Therefore, we investigate the application of self-supervised learning (SSL) methods to effectively utilize the abundant unlabeled images freely available from whole slide images (WSIs) of Pap smear tests. Similarly, we analyze the potential and difficulties of deep learning-based cervical cytology diagnosis and systematically report our findings. Specifically, we unveil the following aspects of deep learning-based Pap smear cytology:

\begin{itemize}
    \item We thoroughly evaluate the ability of models pre-trained with self-supervised learning to learn meaningful visual representations of cytology images for various downstream tasks. In particular, the resulting representations show superior discriminability and generalizability;
    \item Our experiments reveal that representations learned with publicly available single cervical cell datasets, \eg Herlev \cite{jantzen2005pap}, or Sipakmed \cite{8451588}, do not generalize well to different modalities such as images representing multiple cells. To mitigate this issue, we propose a data augmentation strategy tailored for cytology images dubbed \textbf{C}ervical \textbf{C}ell \textbf{C}opy-\textbf{P}asting ($\pname$). Furthermore, we demonstrate the effectiveness of $\pname$ for learning generalizable representations from single-cell datasets;
    \item We experimentally observe that multiple instance learning (MIL), the commonly used strategy for obtaining WSI-level representations and predictions, does not fully exploit the inherent properties of Pap smear cytology slides. Consequently, we introduce a set of simple yet effective modifications, \eg processing only the top-k most suspicious instances, to better align MIL methods with Pap smear test images;
    \item We present a medium-size liquid-based cytology Pap smear test images dataset from HPV-positive women. The slides of this dataset are prepared with the SurePath\textsuperscript{\texttrademark} procedure, which results in a small cell-deposit area. This shortens the time of digitization and yields smaller WSIs files, which is ideal for our telecytologic-based same-day ``test-triage-and-treat'' objective. The presented dataset is particularly challenging as all samples are from HPV-positive women, and negative slides typically portray signs of infections, which complicates the diagnosis. 
\end{itemize}

\section{Related Work}
\label{sec:related_work}
The advent of digital microscopy and the recent evidence that digitally based diagnostic performance is on-par with light microscopy for Papanicolau test slides \cite{kholova2022inter} provides an interesting alternative. The added possibility of incorporating computer-assisted and/or automated diagnostics makes this an even more exciting prospect. Histopathology-based assessment is considered the ``gold standard'' for most cancer-related diagnostics, and histology attracts far more attention from the machine learning community \cite{ABBET2022102473,stegmuller2023scorenet,bozorgtabar2021sood} than cytology. Recently though, cytopathology has gained more traction and recognition, as it offers a non-invasive and inexpensive diagnostic tool suited to resource-constrained countries. Consequently, there have been significant advancements in machine-learning approaches applied to cytology. \par
Most of these advancements focus on cell-level tasks, \eg classification \cite{lin2019fine,albuquerque2021ordinal}, detection \cite{liang2021global,li2019detection} or segmentation \cite{hussain2020shape}. These innovations aim to improve the efficiency and accuracy of cervical cancer screening and other forms of cancer diagnosis from cytological samples. The classification of whole slide Pap smear test images remains significantly less studied, despite its practical application being the most promising. Notable works on the topic include \cite{cheng2021robust}, who combined low- and high-resolution stages for the identification/localization of suspicious lesions and their classification. The high-resolution stage relies on a recurrent neural network (RNN)-based classification model to predict the WSI-level scores. Another study, \cite{wei2021efficient}, leveraged a YOLO-based \cite{redmon2018yolov3} approach to generate cell/tile-level predictions in the first stage and a transformer model for the aggregation and WSI-level classification in the second stage. Similarly, \cite{cao2021novel} proposed the integration of an attention module to detect abnormal cells in large patches and computed the abnormality probability of a given patch as the average of its constituent cells. The WSI-level score is obtained by averaging the abnormality of its patches. Recently, \cite{li2023novel} proposed a three-stage pipeline for lung cancer cytopathological WSIs classification. Their approach integrates a transformer-based model to extract fine-grained lesion features, which are then aggregated into intermediate patch-level features, and coarse-grained features for final WSI-level classification.
 \par
\section{Datasets}
\label{ssec:datasets}
\begin{figure*}[hb]
    \centering
    \includegraphics[width=\textwidth]{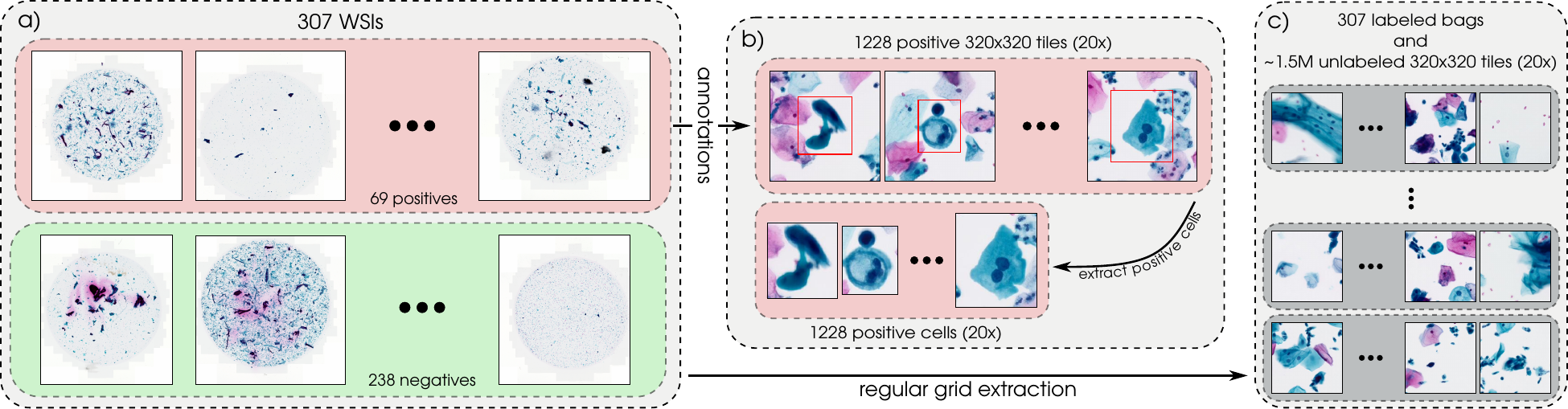}
    \caption{
    \textbf{Overview of the in-house dataset.} a) The $N$ WSIs are labeled as positives ($N_{p}$ samples) or negatives ($N_{n}$ samples). b) 1228 cell-level annotations are used to extract isolated positive cells and $320\times320$ pixels tiles, both at 20$\times$ magnification. c) The tiles of each WSI are extracted based on a regularly spaced grid, yielding $\sim$1.5M unlabeled tiles ($320\times320$ pixels at 20$\times$ magnification) distributed over 307 slide bags.
    }
    \label{fig:dataset}
\end{figure*}
\noindent
{\bf In-house dataset.}
\label{par:inhouse_dataset}
We present our in-house dataset composed of a  cohort of 307 Pap smear slides from HPV-positive patients. The prevalence of cytology-positive slides is approximately $20\%$, translating to 69 positive and 238 negative slides. The preparation of the slides follows the SurePath\textsuperscript{\texttrademark} procedure. This choice is motivated by the overall objective of the telecytologic diagnosis of cervical smears for the triage of HPV-positive women in a resource-limited setting. Indeed, this preparation yields a small cell-deposit area, which shortens the scanning time and reduces the size of the digitized slides. Additionally, the SurePath\textsuperscript{\texttrademark} procedure exists in a manual and low-cost version to further ease its adoption in a low-income setting. Along the same line, the slides are digitized with the \href{https://www.grundium.com/ocus40/}{Grundium Ocus\textsuperscript{\textregistered}40} scanner: a portable and affordable solution. The WSIs are acquired with a 12 megapixels image sensor, a 40x objective, and Z-stacking (3 focal planes spaced by 1$\mu m$).
 \par
After digitization, cell-level annotations are obtained using QuPath~\cite{bankhead2017qupath}, resulting in a total of 1228 annotated positive cells. The annotations are used to create a dataset of 1228 positive cell images and as many positive $320\times320$ pixels tiles (cell+context, see \cref{fig:dataset}) both at $20\times$ magnification (0.50 $\mu m / px$). Alternatively, we create an unlabeled tile dataset by finding the cell-deposit area with standard image processing techniques and subsequently sampling $320\times320$ tiles on a regular grid without overlap, yielding approximately 1.5M images, subdivided into 307 bags encompassing an average of 5200 tiles each. The bags are used for WSI-level classification (see \cref{ssec:m_e_mil}), whereas the individual tiles serve as a substrate for the SSL pre-training (see \cref{par:ssl_pretraining}). We use a stratified 4-fold split approach to partition the slides into training, validation, and test subsets. The slide-level splits are common to all experiments, including the SSL pre-training.

\noindent
{\bf Herlev.}
\label{par:herlev}
A liquid-based cytology Pap smear tests images dataset \cite{jantzen2005pap} encompassing 917 labeled cell images. It contains 242 cytology-negative images of annotated cell types of \textit{superficial squamous epithelial} (NS), \textit{intermediate squamous epithelial} (NI), and \textit{Columnar epithelial} (NC). The 675 cytology-positives images are annotated as lesion cells of \textit{mild squamous non-keratinizing dysplasia} (LD), \textit{moderate squamous non-keratinizing dysplasia} (MD), \textit{severe squamous non-keratinizing dysplasia} (SD), and \textit{squamous cell carcinoma in situ intermediate} (CIS), respectively.

\noindent
{\bf Sipakmed.}
\label{par:sipakmed}
A dataset of cervical squamous cells \cite{8451588} from Pap smear images, which comprises 4049 labeled cell images. The 2411 cytology-negative images are further categorized in \textit{metaplastic} (M), \textit{superficial-intermediate} (SI), and \textit{parabasal} (P). Similarly, images are annotated as \textit{Koilocytotic} (K) and \textit{dyskeratotic} (D) are represented among the 1638  cytology-positive images.

\section{Method \& Experiments}
\label{sec:method_experiments}

\begin{figure*}[ht]
    \centering
    \includegraphics[width=\textwidth]{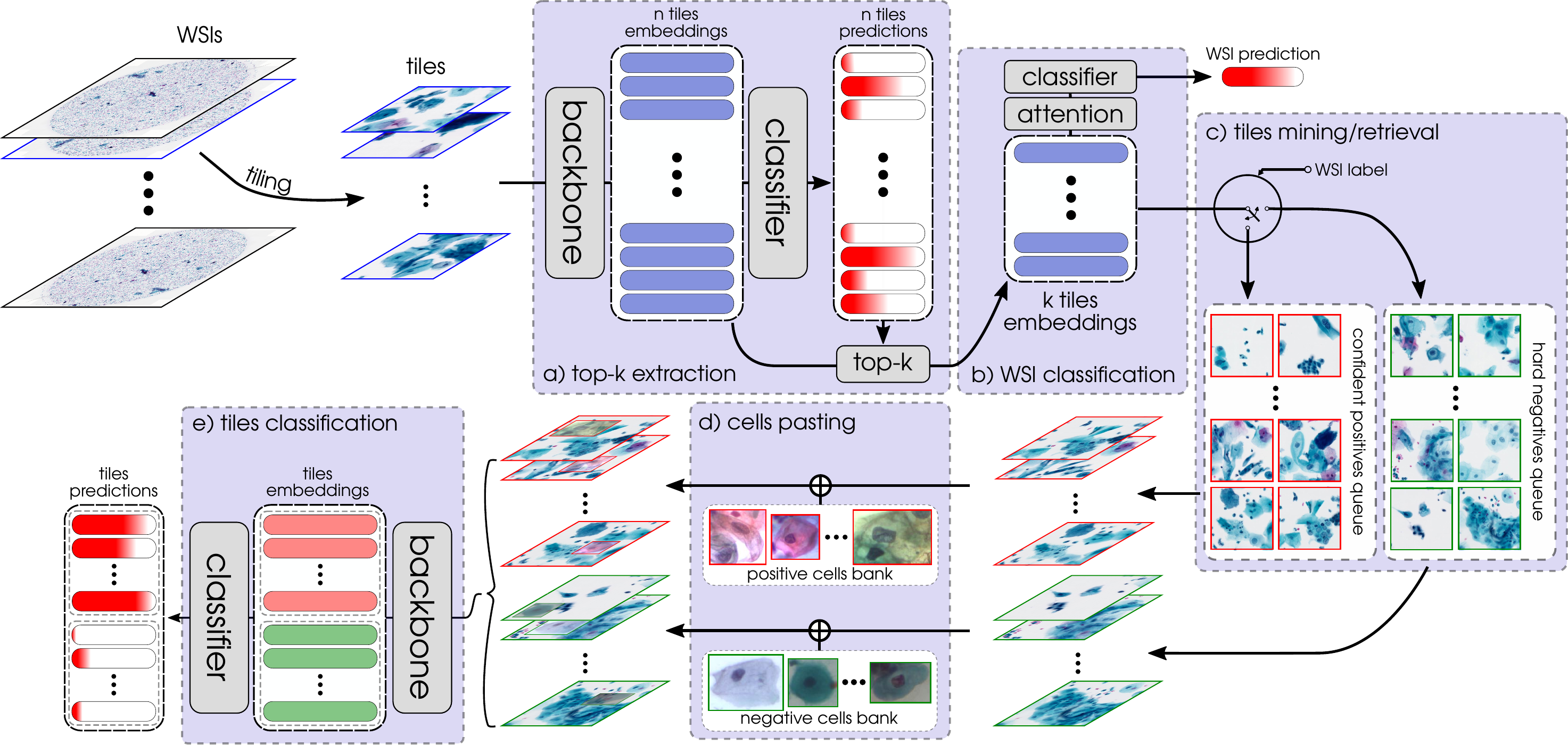}
    \caption{
    \textbf{Overview of the proposed MIL based method for classifying Pap smear WSIs.} a) A positivity score is obtained \textbf{independently} for each tile of the input WSI, and the embeddings of the tiles having the top-k highest scores are extracted. b) The top-k embeddings attend to one another to produce the slide-level representation, where the positivity score is obtained using the \textbf{same classifier} as for the independent tiles predictions. c) The tiles corresponding to the top-k scores are stored as \textit{confident positives} or \textit{hard negatives} queues, depending on the slide-level label. d) Positive and negative cells are pasted upon randomly sampled \textit{confident positives} and \textit{hard negatives}, respectively. e) A score for each pasted tile is obtained using the \textit{same backbone and classifier}. The model is conjointly trained to correctly classify WSIs and pasted tiles.
    }
    \label{fig:pipeline}
\end{figure*}

In \Cref{sec:introduction_new}, we evoke the urge to develop robust machine-learning methods for diagnosing cervical cancer based on Pap smear cytology in a low-resource setting. The emergence of affordable and transportable high-resolution scanners, \eg \href{https://www.grundium.com/ocus40/}{Grundium Ocus\textsuperscript{\textregistered}40}, as well as the existence of low-cost slide preparation procedures (SurePath\textsuperscript{\texttrademark}), make for a particularly ripe ground for this enterprise. 

Nonetheless, collecting extensive and meticulously curated annotated data is time-consuming and expensive. Consequently, we investigate self-supervised learning methods and approaches that require few labeled samples. More precisely, in~\Cref{ssec:m_e_ssl}, we provide empirical evidence that self-supervised learning approaches can be successfully leveraged to learn meaningful representations of multiple-cell images, \ie unlabeled tiles.  Our proposed cell augmentation method $\pname$ is discussed and extensively tested in~\Cref{ssec:m_e_pasting}. Finally, in~\Cref{ssec:m_e_mil}, we provide and discuss simple, yet effective tools tailored to existing MIL methods for cytology Pap smear test WSIs.

\subsection{How well do self-supervised models transfer to cytology images?}
\label{ssec:m_e_ssl}

While self-supervised learning (SSL) methods have gained attention for diverse downstream tasks in histopathology images, these approaches have received little attention for cytology-related tasks due to little evidence of their feasibility on Pap smear cytology images. Most state-of-the-art SSL methods \cite{he2020momentum,zhou2021ibot,caron2020unsupervised,caron2021emerging} for image-level representations learning rely on maximizing the similarity of an image's representation under information preserving transformations. Crucially, one of these transformations is a spatial crop, which is at risk of losing its information-preserving property on cytology images as a consequence of the preparation of the slides that break long-range spatial dependencies. On the contrary, each digitized cytology slide can yield thousands of unlabeled images, which indicates that cytology could potentially be a playground where SSL methods thrive. \par
\noindent
{\bf Self-supervised pre-training.}
\label{par:ssl_pretraining}
We pre-train our models, using $\dino$ \cite{caron2021emerging} as the self-supervised learning framework. This choice is motivated by its strong nearest neighbor classifier capability and excellent performance across different backbones. $\dino$ relies on a pair of Siamese teacher-student networks and a knowledge distillation approach. The underpinning principle of the method is to train the student network to mimic the teacher's output distribution when both models are fed with distinct views of the same input image. DINO leverages both \textit{global views} and \textit{local views}. The former typically spans a larger image region and captures image-level dependencies, while the latter occupies a fraction of the image and yields localized features. By leveraging views at different scales, local-to-global consistency can be distilled from the teacher to the student network. Compared to contrastive learning approaches \citep{chen2020simple,he2020momentum}, self-distillations methods \citep{caron2021emerging,grill2020bootstrap} must explicitly avoid the collapse of the learned representation to trivial solutions. In particular, $\dino$ only updates the teacher network's weights with an exponential moving average (EMA) of those of the student network. Towards the same objective, the entropy of the teacher's output distribution is constrained with sharpening and centering tricks.\par

We experiment with two types of architecture, $\resnet$-50 \cite{he2016deep} and vision transformer (ViT), $\vit$-S/16 \cite{dosovitskiy2020image}, for which we use the recommended parameters available on the official \href{https://github.com/facebookresearch/dino}{repository}. The arguments only differ from the recommendations for the batch size and the number of local crops. The batch size is set to fill the available GPU memory, \ie $\texttt{batch\_size} = 256$ for a $\resnet$-50 and $\texttt{batch\_size} = 192$ for a $\vit$-S/16. We do not use local crops as they can result in ambiguous positive pairs for non-object-centric datasets, as is the case here. For each architecture, we train one model per stratified split (see ~\cref{par:inhouse_dataset}) for 300 epochs; hence we obtain four pre-trained models for each architecture.
\par
In all the following experiments, we compare the quality of the learned visual representations under the above-described setting to the ones obtained under a supervised pre-training on $\imagenet$. For the $\resnet$-50 architecture, we use the weights provided by PyTorch \cite{paszke2019pytorch}, whereas, for the $\vit$-S/16, we rely on the weights of \cite{pmlr-v139-touvron21a} (trained without distillation).\par
\noindent
{\bf Cell-level classification.}
\label{par:cell_knn}
After model pre-training, we probe the quality of the learned features on a cell-level classification task. We opt for a k-NN classifier in limiting the manual intervention to the minimum, thereby obtaining results that reflect the learned representations' quality. We use two publicly available cervical cell Pap smear test datasets: the Herlev dataset \cite{jantzen2005pap} and the Sipakmed dataset \cite{8451588}. These datasets are randomly split in train/validation with a 75/25 partition. We report the mean and standard deviation of the class-wise and weighted $F_{1}$ scores over $4$ independent runs for each pre-trained model, \ie a total of $16$ for the models pre-trained under the self-supervised framework (see~\cref{par:ssl_pretraining}) and $4$ runs for the supervised ones. The number of neighbors $k$ is selected to maximize the weighted $F_{1}$ score. \par

In~\Cref{table:herlev_knn,table:sipakmed_knn}, we observe that despite being pre-trained without any labels and not on isolated cells, the models resulting from $\dino$'s pre-training are on-par or better than the ones pre-trained on $\imagenet$, which are competitive baselines and the \textit{de facto} choice for most practitioners. It further appears that the Herlev dataset is more challenging, especially with fine-grained class labels. However, the representations are good enough to differentiate negative cells from positive ones.
\begin{table*}[ht]
\centering
\caption{
\textbf{Cell-level classification results on Herlev. We report the class-wise and weighted $F_{1}$ scores of a k-NN classifier.} The features are extracted by a $\vit$-S/16 or a $\resnet$-50 pre-trained under a supervised pre-training on ImageNet or a \hl{self-supervised pre-training} on our in-house unlabeled tiles dataset using DINO. The highest mean score for a given class and backbone are highlighted in \textbf{bold}.
}
\footnotesize
\resizebox{1.0\textwidth}{!}{
\begin{tabular}{l c c c c c c c c c c c c c}
\toprule
& & \multicolumn{4}{c}{\textit{positives}} && \multicolumn{3}{c}{\textit{negatives}} && \multicolumn{3}{c}{\textit{average}} \\
\cmidrule{3-6} \cmidrule{8-10} \cmidrule{12-14}
\textit{backbone} & & CIS & LD & MD & SD && NC & NI & NS && positives & negatives & weighted $F_{1}$ \\
\midrule
ResNet-50 & & $55.0 \pm 4.5$ & $64.3 \pm 5.3$ & $\mathbf{48.6 \pm 4.8}$ & $51.3 \pm 3.7$ && $\mathbf{53.8 \pm 4.1}$ & $87.0 \pm 7.5$ & $89.6 \pm 6.3$ && $\mathbf{93.2 \pm 0.6}$ & $\mathbf{80.1 \pm 1.7}$ & $60.2 \pm 3.2$ \\
\rowcolor{cyan!20} ResNet-50 & & $\mathbf{59.0 \pm 7.1}$ & $\mathbf{66.2 \pm 3.7}$ & $\mathbf{48.6 \pm 6.8}$ & $\mathbf{53.0 \pm 4.9}$ && $50.9 \pm 8.0$ & $\mathbf{88.8 \pm 5.2}$ & $\mathbf{92.4 \pm 2.8}$ && $92.3 \pm 1.5$ & $78.6 \pm 3.8$ & $\mathbf{61.7 \pm 3.2}$ \\
\midrule
ViT-S/16 & & $55.1 \pm 5.7$ & $59.5 \pm 3.2$ & $38.6 \pm 8.2$ & $48.7 \pm 3.1$ && $49.2 \pm 4.6$ & $75.8 \pm 5.1$ & $83.2 \pm 7.2$ && $92.4 \pm 1.1$ & $76.8 \pm 2.4$ & $55.2 \pm 2.0$ \\
 \rowcolor{cyan!20} ViT-S/16 & & $\mathbf{62.8 \pm 4.4}$ & $\mathbf{66.8 \pm 4.5}$ & $\mathbf{48.2 \pm 5.1}$ & $\mathbf{53.5 \pm 5.5}$ && $\mathbf{58.8 \pm 7.5}$ & $\mathbf{83.4 \pm 1.9}$ & $\mathbf{89.7 \pm 3.2}$ && $\mathbf{93.1 \pm 0.9}$ & $\mathbf{80.8 \pm 2.9}$ & $\mathbf{62.6 \pm 2.8}$ \\
\bottomrule
\end{tabular}
\label{table:herlev_knn}
}
\end{table*}%

\begin{table*}[ht]
\centering
\caption{
\textbf{Cell-level classification results on Sipakmed. We report the class-wise and weighted $F_{1}$ scores of a k-NN classifier.} The features are extracted by a $\vit$-S/16 or a $\resnet$-50 pre-trained under a supervised pre-training on ImageNet or a \hl{self-supervised pre-training} on our in-house unlabeled tiles dataset using DINO. The highest mean score for a given class and backbone are highlighted in \textbf{bold}.
}
\footnotesize
\begin{tabular}{l c c c c c c c c c c c c}
\toprule
 & & \multicolumn{2}{c}{\textit{positives}} && \multicolumn{3}{c}{\textit{negatives}} && \multicolumn{3}{c}{\textit{average}} \\
\cmidrule{3-4} \cmidrule{6-8}  \cmidrule{10-12}
\textit{backbone} & & D & K && M & P & SI && positives & negatives & weighted $F_{1}$ \\
\midrule
 ResNet-50 & & $\mathbf{94.5 \pm 1.4}$ & $85.0 \pm 0.8$ && $89.2 \pm 1.2$ & $94.4 \pm 1.4$ & $97.2 \pm 0.4$ && $94.5 \pm 0.4$ & $96.4 \pm 0.3$ & $92.1 \pm 0.6$ \\[0.9mm]
\rowcolor{cyan!20} ResNet-50 & & $93.7 \pm 0.9$ & $\mathbf{87.7 \pm 1.7}$ && $\mathbf{91.2 \pm 1.3}$ & $\mathbf{97.2 \pm 0.9}$ & $\mathbf{98.6 \pm 0.6}$ && $\mathbf{95.5 \pm 0.6}$ & $\mathbf{96.9 \pm 0.4}$ & $\mathbf{93.7 \pm 0.7}$  \\ [0.9mm]
\midrule
ViT-S/16 & & $89.6 \pm 1.8$ & $82.6 \pm 2.8$ && $85.2 \pm 2.3$ & $94.2 \pm 0.6$ & $95.3 \pm 0.7$ && $93.4 \pm 1.1$ & $95.5 \pm 0.8$ & $89.3 \pm 1.3$ \\ [0.9mm]
\rowcolor{cyan!20} ViT-S/16 & & $\mathbf{94.8 \pm 0.8}$ & $\mathbf{88.1 \pm 1.3}$ && $\mathbf{91.1 \pm 1.2}$ & $\mathbf{98.0 \pm 0.5}$ & $\mathbf{98.5 \pm 0.3}$ && $\mathbf{95.6 \pm 0.6}$ & $\mathbf{97.0 \pm 0.4}$ & $\mathbf{94.1 \pm 0.4}$\\ [0.9mm]
\bottomrule
\end{tabular}
\label{table:sipakmed_knn}
\end{table*}%

\begin{table}[ht]
\centering
\caption{
\textbf{Tiles-level evaluation results of the frozen models} on the in-house set of tiles. A k-NN classifier is fitted on $75\%$ of the samples and evaluated on the remaining $25\%$. We report the class-wise and weighted $F_{1}$ scores of $4$ independent runs. The features are extracted by a $\vit$-S/16 or a $\resnet$-50 pre-trained under a supervised pre-training on ImageNet or a \hl{self-supervised pre-training} on our internal dataset using DINO. The highest mean score for a given class and backbone is highlighted in \textbf{bold}.
}

\footnotesize
\resizebox{1.0\columnwidth}{!}{
\begin{tabular}{l c c c c }
\toprule
\textit{backbone} && positives & negatives & weighted $F_{1}$\\
\midrule
$\resnet$-50 && $81.6 \pm 0.3$ & $79.1 \pm 0.5$ & $80.4 \pm 0.2$ \\
\rowcolor{cyan!20} $\resnet$-50 && $\mathbf{95.3 \pm 0.6} \:\textcolor{dg}{(+13.7)}$ & $\mathbf{95.1 \pm 0.7} \:\textcolor{dg}{(+16.0)}$ & $\mathbf{95.2 \pm 0.6} \:\textcolor{dg}{(+14.8)}$\\
\bottomrule
$\vit$-S/16 && $77.8 \pm 0.6$ & $72.1 \pm 1.8$ & $74.9 \pm 1.0$ \\
\rowcolor{cyan!20} $\vit$-S/16 && $\mathbf{96.6 \pm 0.4} \:\textcolor{dg}{(+18.8)}$ & $\mathbf{96.4 \pm 0.5} \:\textcolor{dg}{(+24.3)}$ & $\mathbf{96.5 \pm 0.4}  \:\textcolor{dg}{(+21.6)}$ \\
\bottomrule
\end{tabular}
}
\label{table:knn_tiles}
\end{table}%

\noindent
{\bf Tile-level classification.}
\label{par:tile_knn}
The representations learned via self-supervised learning are also evaluated on a tile-level classification task. As for the cell-level classification task, we rely on a k-NN approach. To that end, we prepare a labeled tiles dataset composed of our in-house $1228$ positive tiles (see~\cref{ssec:datasets}) and as many tiles randomly sampled from negative slides. The k-NN classifier is fitted on $75\%$ of the resulting dataset and tested against the remaining $25\%$. We use the same evaluation setting as for the above-described cell-level classification task.

Various conclusions can be drawn from the results depicted in~\Cref{table:knn_tiles}. First, we achieve superior performance for tile-level classification using a simple k-NN classifier and \textit{off-the-shelf} pre-trained models. Nonetheless, a given positive tile typically represents more negative cells than positive ones; hence it could be anticipated that the positive signal of a tile would get overshadowed. Secondly, we observe that the self-supervised pre-training yields a significant boost in performance when the pre-training and target datasets are well aligned. Overall, it is remarkable that the self-supervised pre-trained models of DINO transfer well to cytology images,  while the DINO approach is originally tailored for object-centric datasets. Furthermore, the quality of the classification obtained with a k-NN classifier only seems to imply that the SSL models do not encode multiple cells as a single pattern, as it would not allow for the matching of positive tiles. We postulate that this is a consequence of the random cropping operation. \par

\subsection{\textbf{C}ervical \textbf{C}ell \textbf{C}opy-\textbf{P}asting: $\pname$}
\label{ssec:m_e_pasting}
\begin{figure*}
    \centering
    \includegraphics[width=\textwidth]{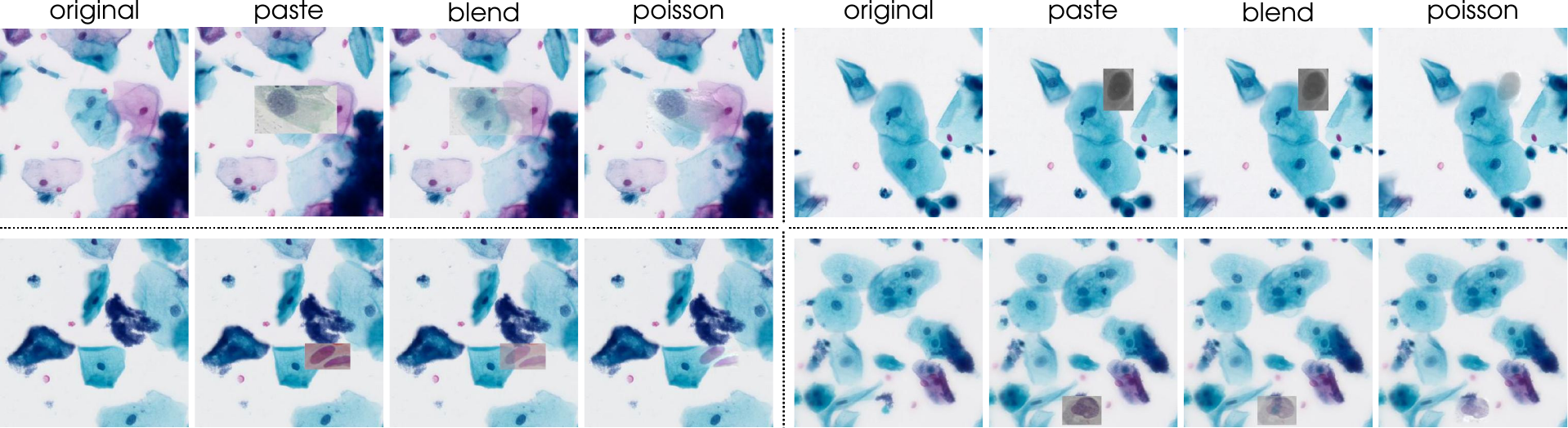}
    \caption{
    Visualization of the different pasting approaches on randomly sampled tiles from the in-house dataset and random pasted cells from both Herlev and Sipakmed datasets.
    }
    \label{fig:pasting_methods}
\end{figure*}
\begin{figure}
    \centering
    \includegraphics[width=\columnwidth]{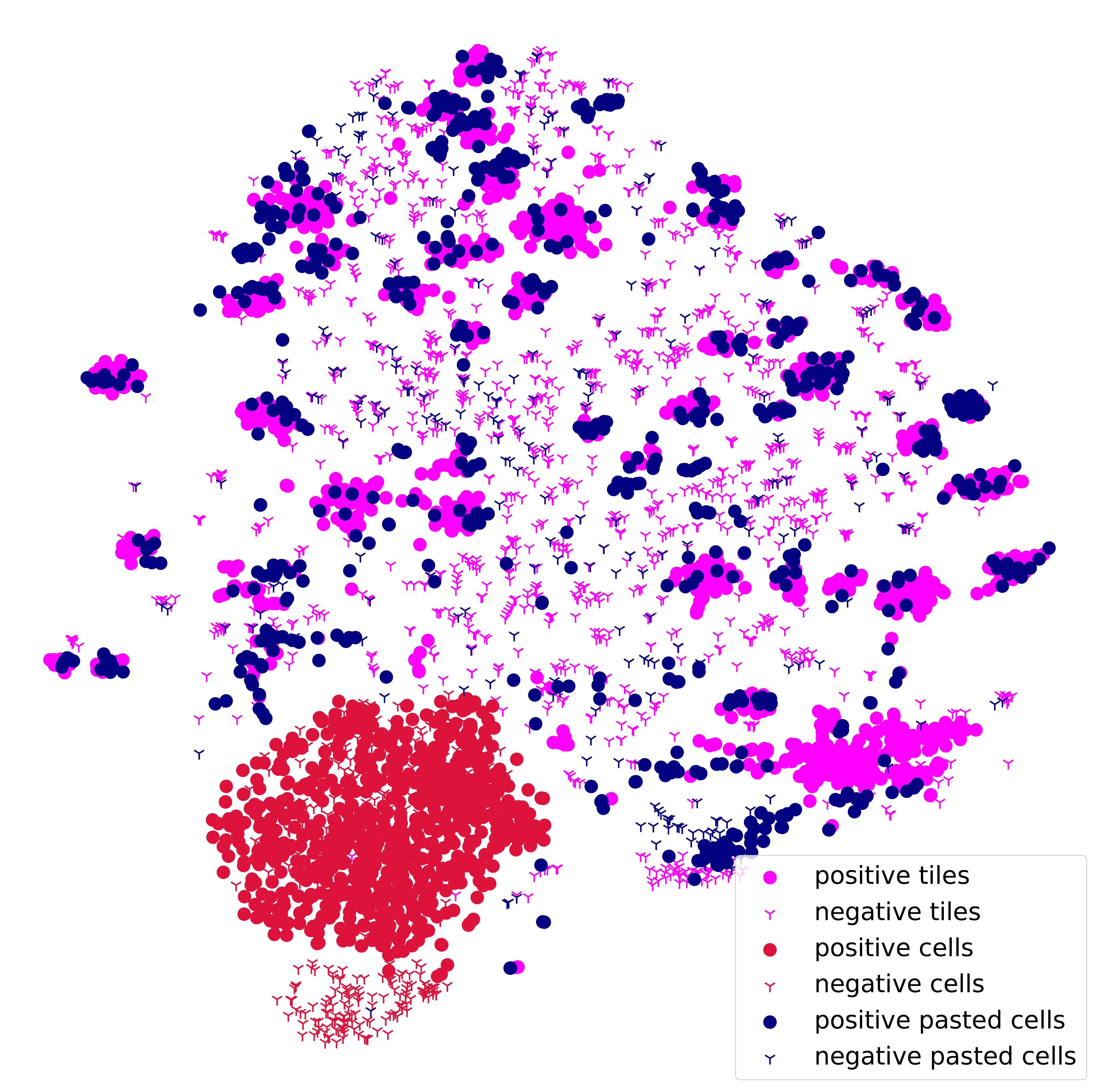}
    \caption{\textbf{The t-SNE projection} obtained from a $\vit$-S/16  encoder of cells from Herlev, our \textit{in-house} labeled tiles, and Herlev cells augmented with $\pname$-\texttt{poisson}.
    }
    \label{fig:herlev_tsne}
\end{figure}
In~\Cref{ssec:m_e_ssl}, we discuss the applicability of self-supervised learning to cytology images and report evidence of its effectiveness. As much as self-supervised learning is an adequate approach that can yield semantically coherent clusters of image representations, it doesn't allow for the labeling of the aforementioned clusters. We investigate if this labeling operation can be performed using publicly available datasets. The major obstacle to achieving this objective is that most public datasets are at the cell level annotations, whereas a lot of cytology tasks, \eg whole slide image classification, require patch/tile level representations and annotations. Consequently, we first show that naively using models trained on cell-level datasets does not transfer well to tile-level downstream tasks. We then propose a simple yet effective method to palliate this issue. \par

\noindent
{\bf Cells to tiles transfer learning.}
\label{par:cells_to_tiles}
 We first evaluate the capability of a classifier trained on open-source cell-level datasets for the tile-level classification at test time. To that end, a k-NN classifier is fitted on the Herlev or Sipakmed datasets using only binary labels, \ie negative or positive, and subsequently evaluated on our \textit{in-house} set of labeled tiles. For each pre-trained model, we report the class-wise $F_{1}$ score averaged over $4$ independent runs, which only use 75\% of the training set each. When the model is pre-trained in a self-supervised manner, the scores are further averaged over the pre-training splits (see~\cref{par:ssl_pretraining}). The number of neighbors $k$ is selected to maximize the $F_{1}$ score of the positive class.\par
 \begin{table}[ht]
\centering
\caption{
\textbf{Transfer learning results from Herlev and Sipakmed to our in-house labeled tiles dataset.} A k-NN classifier is fitted on the binary version of the Herlev (H) and Sipakmed (S) datasets and is evaluated on the in-house set of labeled tiles. The features are extracted by a $\vit$-S/16 or a $\resnet$-50 pre-trained under a supervised pre-training on $\imagenet$ or a \hl{self-supervised pre-training} on our in-house unlabeled tiles dataset using DINO. The highest mean score for a given source dataset and backbone is highlighted in \textbf{bold}.
}

\footnotesize
\resizebox{1.0\columnwidth}{!}{
\begin{tabular}{l c c c c c c c c c c c}
\toprule
&& \multicolumn{2}{c}{\textit{Herlev $\rightarrow$ tiles}} && \multicolumn{2}{c}{\textit{Sipakmed $\rightarrow$ tiles}} \\
\cmidrule{2-4} \cmidrule{6-7}
\textit{backbone} && negatives & positives  && negatives & positives \\ 
\midrule
$\resnet$-50                    && $33.3 \pm 5.7$ & $0.0 \pm 0.0$ && $\mathbf{68.9 \pm 2.9}$ & $0.0 \pm 0.0$ \\ 
\rowcolor{cyan!20} $\resnet$-50 && $\mathbf{63.7 \pm 1.6}$ & $\mathbf{33.8 \pm 5.9}$ && $65.3 \pm 1.2$ & $\mathbf{18.4 \pm 4.8}$ \\
\midrule
$\vit$-S/16                    && $39.5 \pm 3.9$ & $0.0 \pm 0.0$ && $\mathbf{74.8 \pm 2.8}$ & $0.0 \pm 0.0$\\ 
\rowcolor{cyan!20} $\vit$-S/16 && $\mathbf{50.4 \pm 1.4}$ & $\mathbf{41.3 \pm 2.6}$ && $61.3 \pm 1.8$ & $\mathbf{30.2 \pm 6.9}$ \\
\bottomrule
\end{tabular}
}
\label{table:cells_to_tiles_direct}
\end{table}%

The results reported in~\Cref{table:cells_to_tiles_direct} clearly show that a direct transfer learning from cells to tiles with
a k-NN classifier performs poorly. More precisely, it can be observed that the models pre-trained in a supervised cannot detect the discriminant signal from the positive tiles. Nonetheless, we cannot conclude that the failure is a consequence of the shift in modality, \ie single-cell images to multi-cell images, and not due to the small capacity of the classifier, the backbone, or another domain discrepancy between the source and target datasets. Furthermore, the self-supervised pre-trained models generalize better on this task. \par

\noindent
{\bf Cells to tiles transfer learning with pasting.}
\label{subpar:cells_to_tiles_pasting}
The above paragraph reveals the inability of a k-NN classifier to transfer learning from cells to tiles images. To shed some light on the underlying cause of this failure, we repeat the same experiment except that, as a pre-processing step, we use the proposed augmentation \ie we paste all the cells from Herlev or Sipakmed upon randomly sampled tiles from negative slides, referred to as canvases. The label of the pasted cell is attributed to the resulting pasted tile. In this first pasting scenario, we use the most straightforward pasting technique, which is referred to as the \texttt{paste} strategy. \par
\noindent
\texttt{paste:}
\label{sspar:pasting_paste}
The strategy relies on a two-step procedure to paste a cell on a tile: \textit{i)} the pasting location of the cell is uniformly sampled among all the positions that would allow the cell to fit entirely in the tile, and \textit{ii)} the pixels of the tile in the pasting site are replaced by those of the cell.
 \par
\par
\begin{table}[ht]
\centering
\caption{
\textbf{Transfer learning results from Herlev and Sipakmed to our in-house labeled tiles dataset using $\pname$-\texttt{paste}.} A k-NN classifier is fitted on the pasted cells from the Herlev (H) and Sipakmed (S) datasets, and evaluated on the in-house set of positive cells/tiles. The features are extracted by a $\vit$-S/16 or a $\resnet$-50 pre-trained under a supervised pre-training on $\imagenet$ or a \hl{self-supervised pre-training} on our in-house unlabeled tiles dataset using DINO. The highest mean score for a given source dataset and backbone is highlighted in \textbf{bold}.
}
\footnotesize
\resizebox{1.0\columnwidth}{!}{
\begin{tabular}{l c c c c c c c c c c c}
\toprule
&& \multicolumn{2}{c}{\textit{Herlev $\rightarrow$ tiles}} && \multicolumn{2}{c}{\textit{Sipakmed $\rightarrow$ tiles}} \\
\cmidrule{2-4} \cmidrule{6-7}
\textit{backbone} && negatives & positives  && negatives & positives \\ 
\midrule
$\resnet$-50  && $29.8 \pm 2.5$ & $65.0 \pm 0.4$ && $46.0 \pm 1.5$ & $65.3 \pm 0.7$ \\ 
\rowcolor{cyan!20} $\resnet$-50 && $\mathbf{41.0 \pm 1.3}$ & $\mathbf{70.7 \pm 0.5}$ && $\mathbf{63.4 \pm 1.0}$ & $\mathbf{74.5 \pm 0.8}$ \\
\midrule
$\vit$-S/16 && $27.1 \pm 0.8$ & $66.0 \pm 0.6$ && $52.9 \pm 0.8$ & $67.4 \pm 1.3$ \\ 
\rowcolor{cyan!20} $\vit$-S/16 && $\mathbf{54.8 \pm 1.4}$ & $\mathbf{75.3 \pm 0.4}$ && $\mathbf{77.9 \pm 0.5}$ & $\mathbf{83.4 \pm 0.3}$ \\
\bottomrule
\end{tabular}
}
\label{table:cells_to_tiles_pasting}
\end{table}%

As can be seen in \Cref{table:cells_to_tiles_pasting}, the proposed augmentation significantly improves the ability of the classifier to detect positive cells in tiles in spite of the large distribution shift between the cells of Helev/Sipakmed and the ones represented in our \textit{in-house} tiles, the small capacity of the classifier and that tiles resulting from \texttt{paste} do not look natural. The t-SNE~\cite{van2008visualizing} mapping depicted in~\Cref{fig:herlev_tsne} shows that the cells and labeled tiles representation are mapped to different regions of the space.  Conversely, positive cells augmented with $\pname$-\texttt{poisson} appear to be close to groups of positive tiles, reflecting the improved alignment obtained with our augmentation strategy.

\noindent
{\it Pasting technique:}
\label{subpar:pasting_method} \\
\begin{table}[ht]
\centering
\caption{
\textbf{Ablation results on pasting method.} A classifier is trained on cells from Herlev or Sipakmed with $\pname$ and various pasting techniques and subsequently evaluated on the in-house labeled tiles. We report the class-wise and weighted $F_{1}$ scores. The highest mean score for a given source dataset, class, and backbone is in \textbf{bold}. The selected pasting technique is \hl{highlighted}.
}

\footnotesize
\resizebox{1.0\columnwidth}{!}{
\begin{tabular}{l c c c c c c c c c c c c}
\toprule
& && \multicolumn{2}{c}{\textit{Herlev}} && \multicolumn{2}{c}{\textit{Sipakmed}} \\
\cmidrule{3-5} \cmidrule{7-8}
\textit{pasting} & \textit{backbone} && negatives & positives  && negatives & positives \\ 
\midrule
\texttt{paste} & $\resnet$-50  && $76.3 \pm 3.3$ & $76.8 \pm 5.6$ && $\mathbf{74.2 \pm 6.0}$ & $\mathbf{77.2 \pm 5.3}$  \\ 
\texttt{blend} & $\resnet$-50 && $72.9 \pm 3.9$ & $75.2 \pm 6.9$ && $67.0 \pm 1.3$ & $70.0 \pm 6.4$ \\ 
\rowcolor{cyan!20} \texttt{poisson} & $\resnet$-50 && $\mathbf{80.4 \pm 2.2}$ & $\mathbf{77.8 \pm 5.9}$ && $73.3 \pm 4.6$ & $71.3 \pm 2.1$ \\
\midrule
\texttt{paste} & $\vit$-S/16 && $\mathbf{73.2 \pm 7.5}$ & $\mathbf{77.1 \pm 8.9}$ && $45.5 \pm 3.3$ & $70.5 \pm 4.7$ \\ 
\texttt{blend} & $\vit$-S/16  && $67.6 \pm 3.1$ & $67.6 \pm 3.6$ && $51.8 \pm 6.2$ & $69.7 \pm 8.4$  \\ 
\rowcolor{cyan!20} \texttt{poisson} & $\vit$-S/16 && $71.2 \pm 9.4$ & $76.7 \pm 8.5$ && $\mathbf{72.6 \pm 8.3}$ & $\mathbf{77.0 \pm 8.9}$ \\
\bottomrule
\end{tabular}
}
\label{table:pasting_method}
\end{table}%
In~\Cref{table:cells_to_tiles_pasting}, we showed that the proposed pasting method could significantly improve the transferability from open-source single-cell datasets to tiles representing multiple cells. Although the pasting method (\texttt{paste}) used to generate the results depicted in~\Cref{table:cells_to_tiles_pasting} works, it is coarse and doesn't produce natural-looking images. As such, it can result in the model focusing exclusively on the pasted regions throughout training, hence performing poorly at test time. Therefore, we investigate if this scenario occurs and if better alternatives exist. In addition to \texttt{paste}, we test two other alternatives referred to as \texttt{blend} and \texttt{poisson}. Examples of samples obtained with the different pasting methods are depicted in~\Cref{fig:pasting_methods}.
\par

\noindent
\texttt{blend:} The only difference w.r.t. \texttt{paste} is that, instead of replacing the pixels of the canvas with those of the cell, the pixels of the pasting site result from a convex combination of those of the cell and canvas:
\begin{equation}
    x_{\texttt{blend}} = (1 - \lambda_{\text{paste}}) \cdot x_{\text{cell}} + \lambda_{\text{paste}} \cdot x_{\text{canvas}}
\end{equation}
where $\lambda_{\text{paste}}$ is sampled uniformly at random from the interval $[0, 1]$. Due to the transparency of the pasting operation, the resulting images look more natural as it mimics the effect of overlapping cells and the border of the cell image is less visible.
 \par
\noindent
\texttt{poisson:}
The main pitfall of the \texttt{blend} strategy is that it can only conceal the boundaries of the pasting site by concealing the cell, which is undesirable. Poisson blending~\cite{perez2003poisson} was precisely proposed to mitigate that issue. Indeed, the blending operation is formulated as an optimization problem, which aims at computing the values of the pixels in the pasting site to preserve the gradients of the source/cell image while matching the pixel intensities of the target/canvas image at the boundaries.
 \par
We train a linear classifier on top of the pre-trained models with different pasting operations. In this experiment, $1000$ unlabeled tiles are used as canvases for each class (negative/positive), and labeled tiles are obtained online by pasting a randomly selected labeled cell upon one of the canvases. Noteworthy that positive cells are pasted upon unlabeled tiles from positive slides and reciprocally for negative cells. After training, the classifier is evaluated on the \textit{in-house} labeled tiles. We report the class-wise $F_{1}$ score averaged over $4$ independent runs per pre-trained weights. The scores are further averaged over the pre-training splits (see~\cref{par:ssl_pretraining}). For this experiment (and the ones that follow), we only use models pre-trained in a self-supervised manner as they have shown to be on par or better than their supervised counterparts. 
 \par
\Cref{table:pasting_method} shows that the \texttt{blend} approach yields worsen results compared to \texttt{paste}. We postulate that this is a consequence of $\lambda_{\text{paste}}$ either being too low and the resulting images not looking more natural than the ones produced with \texttt{paste}, or it being too high and the pasted content being barely visible. On the contrary, we observe that the \texttt{poisson} technique performs similarly to \texttt{paste} for all backbone/dataset combinations, except for the $\vit$-S/16 + Sipakmed scenario, in which case it is the only pasting technique that yields decent results for the classification of negative tiles.

\noindent
{\it Pasting probability:}
\label{subpar:pasting_probability}
\begin{table}[ht]
\centering
\caption{
\textbf{Ablation experiments for pasting probability.} A classifier is trained on cells from Herlev or Sipakmed with various probabilities of applying $\pname$-\texttt{poisson} on \textcolor{dg}{negative (-)} and \textcolor{red}{positive (+)} tiles. The classifier is then evaluated on the in-house labeled tiles. We report the class-wise and weighted $F_{1}$ scores. The highest mean score for a given source dataset, class, and backbone is in \textbf{bold}. The selected pasting method is \hl{highlighted}.
}
\footnotesize
\resizebox{1.0\columnwidth}{!}{
\begin{tabular}{c c  c c c c c c c c c c}
\toprule
& & \multicolumn{2}{c}{\textit{Herlev $\rightarrow$ tiles}} && \multicolumn{2}{c}{\textit{Sipakmed $\rightarrow$ tiles}} \\
\cmidrule{3-4} \cmidrule{6-7}
\textit{pasting [\%] (\textcolor{dg}{-} | \textcolor{red}{+})} & \textit{backbone} & negatives & positives  && negatives & positives \\ 
\midrule
 \:\:\: \textcolor{dg}{0} | \textcolor{red}{100} & $\resnet$-50 & $76.6 \pm 3.3$ & $71.5 \pm 8.2$ && $83.3 \pm 5.5$ & $81.3 \pm 8.0$  \\
\rowcolor{cyan!20}\: \textcolor{dg}{50}  | \textcolor{red}{100} & $\resnet$-50 & $\mathbf{81.6 \pm 2.7}$ & $75.0 \pm 5.2$ && $\mathbf{84.2 \pm 4.2}$ & $\mathbf{82.2 \pm 3.9}$ \\
\textcolor{dg}{100} | \textcolor{red}{100} & $\resnet$-50 & $80.4 \pm 2.2$ & $\mathbf{77.8 \pm 5.9}$ && $73.3 \pm 4.6$ & $71.3 \pm 2.1$ \\
\midrule
 \:\:\: \textcolor{dg}{0} | \textcolor{red}{100} & $\vit$-S/16 & $78.4 \pm 4.8$ & $71.7 \pm 8.7$ && $25.2 \pm 16.0$ & $65.2 \pm 8.5$\\
\rowcolor{cyan!20} \: \textcolor{dg}{50}  | \textcolor{red}{100} & $\vit$-S/16 & $\mathbf{83.1 \pm 2.3}$ & $\mathbf{81.4 \pm 2.1}$ && $\mathbf{80.4 \pm 4.7}$ & $75.7 \pm 11.5$ \\
\textcolor{dg}{100} | \textcolor{red}{100} & $\vit$-S/16 & $71.2 \pm 9.4$ & $76.7 \pm 8.5$ && $72.6 \pm 8.3$ & $\mathbf{77.0 \pm 8.9}$ \\
\bottomrule
\end{tabular}
}
\label{table:pasting_probability}
\end{table}%
So far, we have applied the pasting operation in a perfectly symmetric manner, \ie it is systematically applied independently of the cell's label and that of the slide from which the canvas is extracted. Nonetheless, our setting is inherently asymmetric: on one side, we know with certainty that tiles extracted from negative slides are all negatives; on the other side, little can be said with regard to the label of tiles extracted from positive slides. Furthermore, by systematically using $\pname$, we are encouraging the model to only attend to the pasting site which is undesirable. We propose to exploit the asymmetry of the setting and not systematically use $\pname$ on negative tiles. This further allows the model to learn from real negative examples without the risk of feeding mislabeled samples to the model. Therefore, we replicate the experiment of~\Cref{table:pasting_method}, but this time, $\pname$-\texttt{poisson} is applied on the unlabeled tiles from negative slides with a given probability (see~\cref{table:pasting_probability}).  \par

Although never applying $\pname$ to negative tiles is the scenario in which the model processes the most realistic samples, we observe in~\Cref{table:pasting_probability} that it can be harmful. This observation is unsurprising, considering that in that setting, the positive label is perfectly correlated with the pasting operation. It is in fact surprising that it does not perform even worse. We argue that this is in part due to the ability of $\pname$-\texttt{poisson} to fool the model. On the contrary, the models trained using $\pname$ with a probability of $0.5$ seems to perform favorably compared to the ones using it systematically. Noteworthy that in that setting, the positive label is correlated with the action of pasting.

\noindent
{\it How many canvases are required?:}
\label{subpar:number_canvases}
To answer this question, we repeat the experiment of~\Cref{table:pasting_method}, with a $0.5$ probability of applying $\pname$-\texttt{poisson} and a varying number of canvases per class. \par
\begin{figure*}
    \centering
    \includegraphics[width=\textwidth]{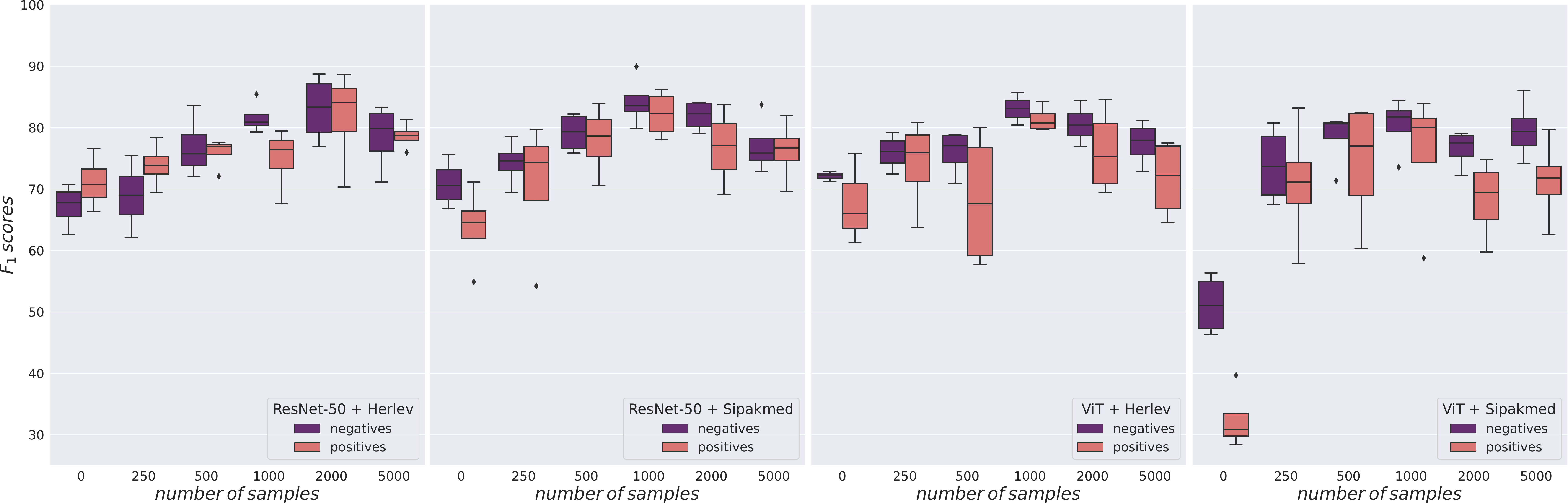}
    \caption{
    \textbf{Box plots depicting the class-wise $F_{1}$ scores against the number of unlabeled tiles used as canvases for the pasting augmentation.} The performance achieved without the proposed augmentation can be observed at the zero of the x-axes.
    }
    \label{fig:boxplot}
\end{figure*}

\Cref{fig:boxplot} depicts the class-wise $F_{1}$ scores for each available backbone/dataset combination. It appears clear that, up until $\approx 2000$ canvases, increasing the number of canvases favorably impacts the classifier's performance. After that point, the model tends to overfit the pasted cells, which occur more often and independently of the canvases, which translates to a decreased downstream performance.

\noindent
{\it $\pname$ results:}
\label{subpar:pasting_results}
\begin{table}[ht]
\centering
\caption{
\textbf{Evaluation results of the cells-pasting augmentation method} with transfer learning from Herlev or Sipakmed to our in-house tiles dataset. A classifier is trained on the cells dataset without and \hl{with $\pname$-\texttt{poisson}}. We report the class-wise and weighted $F_{1}$ scores. The highest mean score for a given backbone, class, and source dataset is highlighted in \textbf{bold}.
}

\footnotesize
\resizebox{1.0\columnwidth}{!}{
\begin{tabular}{ c c c c c c c c c c}
\toprule
&& \multicolumn{2}{c}{\textit{Herlev $\rightarrow$ in-house tiles}} && \multicolumn{2}{c}{\textit{Sipakmed $\rightarrow$ in-house tiles}} \\
\cmidrule{3-4} \cmidrule{6-7}
\textit{backbone} && negatives & positives  && negatives & positives \\ 
\midrule
$\resnet$-50 && $67.2 \pm 3.5$ & $71.1 \pm 4.4$ && $70.9 \pm 3.9$ & $63.8 \pm 6.7$  \\ 
\rowcolor{cyan!20}$\resnet$-50  && $\mathbf{83.1 \pm 5.5} \: \textcolor{dg}{(+15.9)}$ & $\mathbf{81.8 \pm 8.1} \: \textcolor{dg}{(+10.7)}$ && $\mathbf{84.2 \pm 4.2} \: \textcolor{dg}{(+13.3)}$ & $\mathbf{82.2 \pm 3.2} \: \textcolor{dg}{(+18.4)}$ \\ 
\midrule
 $\vit$-S/16  && $72.2 \pm 0.8$ & $67.7 \pm 7.4$ && $52.2 \pm 5.0$ & $32.4 \pm 5.0$ \\ 
\rowcolor{cyan!20} $\vit$-S/16 && $\mathbf{83.1 \pm 2.3} \: \textcolor{dg}{(+10.9)}$ & $\mathbf{81.4 \pm 2.1} \: \textcolor{dg}{(+13.7)}$ && $\mathbf{80.4 \pm 4.7}  \: \textcolor{dg}{(+28.2)}$ & $\mathbf{75.7 \pm 11.5} \: \textcolor{dg}{(+43.3)}$ \\ 
\bottomrule
\end{tabular}
}
\label{table:tiles_classification_pasting}
\end{table}%
Our extended experiments reveal that $\pname$ offers a well-grounded augmentation strategy to bridge the gap between publicly available single-cell and unlabeled tiles datasets. We further show that the proposed augmentation yields significant improvement compared to the approach of naively transferring from a classifier trained on single-cell datasets. In~\Cref{table:tiles_classification_pasting}, we also show that our approach outperforms the naive transferring methods by a large margin with a classifier trained with $\pname$-\texttt{poisson}, a pasting probability of 0.5, and the optimal number of canvases (see~\cref{fig:boxplot}).

\subsection{Aligning MIL to cytology images}
\label{ssec:m_e_mil}
\begin{table}[ht]
\centering
\caption{
\textbf{Evaluation of the MIL-based methods before and after adding our augmentation $\pname$ and \textit{top-k selection} strategy} for Pap-smear test WSIs classification on our \textit{in-house} dataset..}

\footnotesize
\resizebox{\columnwidth}{!}{
\begin{tabular}{c c c c c c c c c c}
\toprule
&&&&& \multicolumn{2}{c}{\textit{AuC scores}} \\
\cmidrule{6-7}
\textit{top-k} | $\pname$ & \textit{method} & \textit{backbone} & $\lambda_{\text{loc}}$ && slide-level & tile-level \\
\midrule
\xk \;|\;   \xk & AbMIL & $\vit$-S/16 & - && $59.0 \pm 11.2$ & $65.9 \pm 11.6$  \\
\ck \;|\; \xk & AbMIL & $\vit$-S/16 & - && $\mathbf{76.8 \pm 3.3}$ & $\mathbf{86.9 \pm 2.2}$  \\
\rowcolor{cyan!20} \ck \;|\; \ck & AbMIL & $\vit$-S/16 & 0.1 && $76.5 \pm 2.9 \:\textcolor{dg}{(+17.5)}$ & $86.1 \pm 2.2 \:\textcolor{dg}{(+20.2)}$  \\
\arrayrulecolor{black!30}\midrule[0.5pt]
\xk \;|\; \xk & AbMIL & $\resnet$-50 & - && $61.1 \pm 11.2$ & $61.6 \pm 12.5$ \\
\ck \;|\; \xk & AbMIL & $\resnet$-50 & - && $70.9 \pm 8.0$ & $74.6 \pm 20.8$ \\
\rowcolor{cyan!20} \ck \;|\; \ck & AbMIL & $\resnet$-50 & 1.0 && $\mathbf{72.8 \pm 2.2} \:\textcolor{dg}{(+11.7)}$ & $\mathbf{80.6 \pm 4.3} \:\textcolor{dg}{(+19.0)}$   \\
\arrayrulecolor{black}\midrule
\xk \;|\; \xk & TransMIL & $\vit$-S/16 & - && $58.1 \pm 5.4$ & $53.3 \pm 9.9$ \\
\ck \;|\; \xk & TransMIL & $\vit$-S/16 & - && $59.8 \pm 14.6$ & $53.7 \pm 12.4$ \\
 \rowcolor{cyan!20} \ck \;|\; \ck & TransMIL & $\vit$-S/16 & 0.1 && $\mathbf{72.1 \pm 8.4} \:\textcolor{dg}{(+14.0)}$ & $\mathbf{67.7 \pm 11.8} \:\textcolor{dg}{(+14.4)}$ \\
\arrayrulecolor{black!30}\midrule[0.5pt]
\xk \;|\; \xk & TransMIL & $\resnet$-50 & - && $49.9 \pm 5.2$ & $46.7 \pm 10.7$  \\
\ck \;|\; \xk & TransMIL & $\resnet$-50 & - && $46.1 \pm 8.7$ & $50.1 \pm 12.9$  \\
\rowcolor{cyan!20} \ck \;|\; \ck & TransMIL & $\resnet$-50 & 1.0  && $\mathbf{69.4 \pm 14.5} \:\textcolor{dg}{(+19.5)}$ & $\mathbf{71.6 \pm 15.1} \:\textcolor{dg}{(+24.9)}$ \\
\arrayrulecolor{black}\midrule
\xk \;|\; \xk & CLAM & $\vit$-S/16 & - && $61.3 \pm 6.2$ & $69.4 \pm 8.9$ \\
\ck \;|\; \xk & CLAM & $\vit$-S/16 & - && $73.8 \pm 4.4$& $\mathbf{84.1 \pm 3.6}$ \\
\rowcolor{cyan!20}  \ck \;|\; \ck & CLAM & $\vit$-S/16 & 0.5 && $\mathbf{74.8 \pm 3.0} \:\textcolor{dg}{(+13.5)}$ & $77.0 \pm 2.9 \:\textcolor{dg}{(+7.6)}$ \\
\arrayrulecolor{black!30}\midrule[0.5pt]
\xk \;|\; \xk & CLAM & $\resnet$-50 & - && $64.3 \pm 11.1$ & $61.4 \pm 11.3$ \\
\ck \;|\; \xk & CLAM & $\resnet$-50 & - && $68.8 \pm 14.4$ & $68.4 \pm 7.2$ \\
\rowcolor{cyan!20} \ck \;|\; \ck & CLAM & $\resnet$-50 & 1.0 && $\mathbf{77.5 \pm 3.4} \:\textcolor{dg}{(+13.2)}$ & $\mathbf{79.0 \pm 4.0} \:\textcolor{dg}{(+17.6)}$ \\
\arrayrulecolor{black}\bottomrule
\end{tabular}
}
\label{table:mil_pasting}
\end{table}%
In~\Cref{ssec:m_e_ssl,ssec:m_e_pasting}, we showcase the benefits of self-supervised learning for cytology diagnostics and propose an augmentation $\pname$ strategy to make the most out of publicly available single-cell datasets. Combined together, this offers the opportunity to design Pap smear WSIs classification modules requiring few labels. More precisely, we harness the power of self-supervised learning and our augmentation strategy $\pname$ to better align well-established MIL methods for Pap smear WSIs classification.

\noindent
{\bf Problem formulation.}
\label{par:problem_formulation}
As a primer, we briefly revisit the underlying concepts and assumptions of the multiple instance learning framework. In a binary MIL setting, the objective is to correctly predict the label $\blabel \in \{ 0,1\}$ of an input bag of instances $\bag = \{ \instance_{1}, \dots, \instance_{n}\}$, where $n$ is allowed to vary from one bag to the other. The instance-level labels $\{ \ilabel_{i} \}_{i=1}^{n} \in \{0, 1 \}$ are assumed to exist but to be unknown throughout the training phase. As such, the MIL objective can be formulated as the detection of positive instances ($y=1$) within the bags, i.e.:
\begin{equation}
\label{eq:bag_label}
    \blabel=\begin{cases}
    1, & \text{iff } \sum_{i} \ilabel_{i} > 0, \\
    0, & \text{otherwise}.
  \end{cases}
\end{equation}
As pointed out in $\abmil$ \cite{ilse2018attention}, the above bag labeling function is permutation invariant w.r.t. the instance labels, hence so must be the predictions $\hat{\blabel} = S(\bag)$, where $S$ is the bag scoring function. In the context of cytology, WSIs are the bags and their constituent tiles are the instances. One can observe that the permutation invariance assumption is particularly well-grounded in that setting. Indeed, the overall diagnosis is based on the presence of abnormal cells within the entire slide rather than the specific arrangement or order of those cells. Furthermore, as a consequence of the slide preparation, the arrangement of the cells on the slides exhibits little to no ordering or positional dependency.
 \par
In most MIL methods, the slide-level representation $\mathbf{z}$ is obtained as a weighted sum/convex combination of the $n$ instance-level representations $H = \{ \mathbf{h}_{1}, \dots, \mathbf{h}_{n}\}$:
\begin{equation}
    \label{eq:weighted_sum}
    \mathbf{z} = \sum_{i} \alpha_{i} \mathbf{h}_{i}
\end{equation}
where $\alpha_{i}$ is a scalar that modulates the contribution of the $i^{th}$ instance to the overall representation. The slide's score is obtained by feeding the slide-level representation to a classifier $g$:
\begin{equation}
    \label{eq:slide_score}
    \hat{\blabel} = g \left(\mathbf{z} \right)
\end{equation}
As the instance-level representations $\mathbf{h}_{i}$ and the slide-level representations $\mathbf{z}$ span the same space, we argue that instance-level predictions can be obtained with the same classifier:
\begin{equation}
    \label{eq:instance_score}
    \hat{\ilabel}_{i} = g \left( \mathbf{h}_{i}\right)
\end{equation}

\noindent
{\bf MIL experimental setup.}
\label{par:mil_exp_setup}
We experiment with $3$ different MIL methods, namely AbMIL \cite{ilse2018attention}, TransMIL \cite{shao2021transmil}, and CLAM \cite{lu2021data}. We directly use their official implementations. We remove the positional encoding from TransMIL as it brings little information in our setting, as discussed above. Each MIL method is tested with both types of backbones (ViT-S/16 and ResNet-50), which are initialized with the weights obtained from DINO's pre-training~\cite{caron2021emerging}. In all experiments, the weights of the backbone are kept frozen. Considering the availability of 4 pre-training weights per backbone (see~\cref{ssec:m_e_ssl}), the first one is used to determine the best hyperparameters, and the three remaining ones are reserved for evaluation purposes. For each setting, we report the average and standard deviation of the slide-level and instance-level AUC scores. The instance-level score is computed using our \textit{in-house} positive tiles and randomly sampled tiles from negative slides (both extracted from the test tiles). \par

\noindent
{\bf Results discussion for MIL-based methods.}
\label{par:mil_results}
In the first scenario, we experiment with the MIL methods using their default implementations. It can be observed in~\Cref{table:mil_pasting} that this setting is suboptimal for all MIL method/backbone combinations. This is intriguing, considering that the backbone demonstrated strong performances (see~\cref{table:knn_tiles}) at the tile level and that the chosen MIL methods are well-established baselines. We argue that this is a consequence of the particularity of Pap smear test images. Indeed, features correlated with negativity are present almost everywhere, even in positive tiles, and features correlated with positivity are scarce. Together, this makes for a particularly challenging setting to capture the positive signal in the slide-level representation, $\mathbf{z}$. \par

\noindent
\textit{Top-k selection:}
\label{subpar:mil_topk}
To mitigate the aforementioned issue, we propose only processing the top-k most suspicious tiles in each slide using the same backbone and classifier as for the slide-level predictions. Noteworthy that as the backbone is frozen, the tiles' representations can be pre-computed, which makes the identification of the top-k tiles not compute-intensive. In all following experiments, we use $k=8$ top-k tiles and a $\texttt{batch\_size} = 16$.
\par
\Cref{table:mil_pasting} shows that adding a top-k module yields significant improvements for all MIL methods except for TransMIL. When using the top-k is beneficial for the slide-level predictions, we observe that it also benefits the tiles-level predictions, which is unsurprising considering that the slide-level representation $\mathbf{z}$ is most likely closer to that of tiles when it results from a weighted-sum over $8$ tiles representations than over the entire bag (see~\cref{eq:weighted_sum}). Along the same line,  the poor tile-level performance of TransMIL is a potential explanation for its ineffectiveness at the slide level. Indeed, if the model cannot detect the positive tiles, the overall representation does not reflect the nature of the slide well. \par

\noindent
\textit{Tile-level objective:}
\label{subpar:mil_tile_objective}
To improve the ability of the model to identify suspicious tiles, we propose to integrate a tile-level loss into the overall training objective:
\begin{equation}
    \label{eq:loc_loss}
    \mathcal{L} = \mathcal{L}_{\text{slide}} + \lambda_{\text{tile}} \cdot \mathcal{L}_{\text{tile}}
\end{equation}
where $\lambda_{\text{tile}}$ denotes the tile loss coefficient. We use a $\texttt{batch\_size}$ of $8$ for the tiles and the optimal value of $\lambda_{\text{tile}}$ is determined independently for each backbone/method combination. Moreover, since we aim for a method using only slide-level labels, we explore the possibility of benefiting from $\pname$. As depicted in~\Cref{fig:pipeline}, \textit{hard negative} and \textit{confident positive} tiles are collected throughout training and used as canvases where negative and positive cells can be pasted upon, respectively. We refer to \textit{hard negatives}/\textit{confident positives} as the $10$ tiles having the highest positivity score in each negative/positive slide, respectively. The method used for pasting is $\pname$-\texttt{poisson}, and we rely on cells from both Herlev and Sipakmed. \par
As shown in~\Cref{table:mil_pasting}, incorporating a localized objective alongside $\pname$ yields significant improvements in the tile-level predictions of TransMIL. Consequently, this facilitates the detection of suspicious tiles and ultimately enhances the accuracy of slide-level predictions. It is worth noting that $\pname$ is not exclusively advantageous for TransMIL, as it proves to be beneficial at the slide level in all scenarios except for $\vit$-S/16 + AbMIL. We posit that a tile-level loss is implicitly enforced when employing a top-k selection approach. In other words, the representation of the top-k selected tiles is encouraged to be aligned with the slide label. Hence, when the top-k selection is accurate, $\pname$ becomes less relevant as ``real'' labeled tiles are available. Nonetheless, this scenario seldom occurs, especially when the backbone is a $\resnet$-50, whose features have 2048 dimensions ($>5\times$ more than for a $\vit$-S/16), which increases the number of parameters of the MIL module. We argue that this explains why settings relying on a $\resnet$-50  as backbone tend to benefit more from $\pname$.

\section{Conclusion}
\label{sec:conclusion}
This paper discussed the potential of deep learning-based telecytology to effectively reduce cervical cancer-related mortality in low-and-middle-income countries by mitigating the need for trained cytopathologists on the triage site. To support this objective, we have collected a medium-sized dataset of Pap smear test images using SurePath\textsuperscript{\texttrademark} preparation, which exists in a manual and low-cost version, and digitized the slides using a Grundium Ocus\textsuperscript{\textregistered}40 scanner, which fulfills our requirements of transportability and affordability. \par

Our experimental findings highlight the successful application of self-supervised learning to reduce the annotation burden, with the resulting representations outperforming \textit{off-the-shelf} pre-trained models across various downstream tasks. Additionally, we have introduced $\pname$, an augmentation strategy, which effectively transfers knowledge from open-source and single-cell datasets to unlabeled tiles. $\pname$ proves to be beneficial not only for tile-level classification but also for slide-level classification. Regarding the WSIs classification, our experimental findings reveal that MIL methods may overlook crucial characteristics of Pap smear images, which can be accounted for by introducing simple modifications that prove to be beneficial. Overall, classifying Pap smear WSIs relying solely on slide-level labels remains challenging, particularly in our scenario where all samples are from HPV-positive women, which adds an additional layer of complexity.
 \par
\noindent
{\bf Limitations.}
Our experiments are conducted on only one self-supervised learning method, namely DINO \cite{caron2021emerging} due to its strong performance on the k-NN evaluation benchmark and compatibility with various backbones. We argue that the main reason why SSL methods could be inadequate for cytology images is that the objective might enforce consistency between semantically unrelated views. Nonetheless, this potential pitfall results from the spatial cropping strategy, which is common to most self-distillation and contrastive methods. Our conclusions based on DINO are likely also applicable to other methods. Alternatively, larger vision transformer backbones, \eg ViT-B/16, would be worth investigating, yet lighter architectures, such as ViT-S/16 and ResNet-50, remain better suited in the low-data regime.



\clearpage

\bibliographystyle{cas-model2-names}

\bibliography{cas-refs}

\begin{thebibliography}{38}
\expandafter\ifx\csname natexlab\endcsname\relax\def\natexlab#1{#1}\fi
\providecommand{\url}[1]{\texttt{#1}}
\providecommand{\href}[2]{#2}
\providecommand{\path}[1]{#1}
\providecommand{\DOIprefix}{doi:}
\providecommand{\ArXivprefix}{arXiv:}
\providecommand{\URLprefix}{URL: }
\providecommand{\Pubmedprefix}{pmid:}
\providecommand{\doi}[1]{\href{http://dx.doi.org/#1}{\path{#1}}}
\providecommand{\Pubmed}[1]{\href{pmid:#1}{\path{#1}}}
\providecommand{\bibinfo}[2]{#2}
\ifx\xfnm\relax \def\xfnm[#1]{\unskip,\space#1}\fi
\bibitem[{Abbet et~al.(2022)Abbet, Studer, Fischer, Dawson, Zlobec, Bozorgtabar
  and Thiran}]{ABBET2022102473}
\bibinfo{author}{Abbet, C.}, \bibinfo{author}{Studer, L.},
  \bibinfo{author}{Fischer, A.}, \bibinfo{author}{Dawson, H.},
  \bibinfo{author}{Zlobec, I.}, \bibinfo{author}{Bozorgtabar, B.},
  \bibinfo{author}{Thiran, J.P.}, \bibinfo{year}{2022}.
\newblock \bibinfo{title}{Self-rule to multi-adapt: Generalized multi-source
  feature learning using unsupervised domain adaptation for colorectal cancer
  tissue detection}.
\newblock \bibinfo{journal}{Medical Image Analysis} \bibinfo{volume}{79},
  \bibinfo{pages}{102473}.
\newblock \URLprefix
  \url{https://www.sciencedirect.com/science/article/pii/S1361841522001207},
  \DOIprefix\doi{https://doi.org/10.1016/j.media.2022.102473}.
\bibitem[{Albuquerque et~al.(2021)Albuquerque, Cruz and
  Cardoso}]{albuquerque2021ordinal}
\bibinfo{author}{Albuquerque, T.}, \bibinfo{author}{Cruz, R.},
  \bibinfo{author}{Cardoso, J.S.}, \bibinfo{year}{2021}.
\newblock \bibinfo{title}{Ordinal losses for classification of cervical cancer
  risk}.
\newblock \bibinfo{journal}{PeerJ Computer Science} \bibinfo{volume}{7},
  \bibinfo{pages}{e457}.
\bibitem[{Bankhead et~al.(2017)Bankhead, Loughrey, Fern{\'a}ndez, Dombrowski,
  McArt, Dunne, McQuaid, Gray, Murray, Coleman et~al.}]{bankhead2017qupath}
\bibinfo{author}{Bankhead, P.}, \bibinfo{author}{Loughrey, M.B.},
  \bibinfo{author}{Fern{\'a}ndez, J.A.}, \bibinfo{author}{Dombrowski, Y.},
  \bibinfo{author}{McArt, D.G.}, \bibinfo{author}{Dunne, P.D.},
  \bibinfo{author}{McQuaid, S.}, \bibinfo{author}{Gray, R.T.},
  \bibinfo{author}{Murray, L.J.}, \bibinfo{author}{Coleman, H.G.}, et~al.,
  \bibinfo{year}{2017}.
\newblock \bibinfo{title}{Qupath: Open source software for digital pathology
  image analysis}.
\newblock \bibinfo{journal}{Scientific reports} \bibinfo{volume}{7},
  \bibinfo{pages}{1--7}.
\bibitem[{Bozorgtabar et~al.(2021)Bozorgtabar, Vray, Mahapatra and
  Thiran}]{bozorgtabar2021sood}
\bibinfo{author}{Bozorgtabar, B.}, \bibinfo{author}{Vray, G.},
  \bibinfo{author}{Mahapatra, D.}, \bibinfo{author}{Thiran, J.P.},
  \bibinfo{year}{2021}.
\newblock \bibinfo{title}{Sood: Self-supervised out-of-distribution detection
  under domain shift for multi-class colorectal cancer tissue types}, in:
  \bibinfo{booktitle}{Proceedings of the IEEE/CVF International Conference on
  Computer Vision}, pp. \bibinfo{pages}{3324--3333}.
\bibitem[{Cao et~al.(2021)Cao, Yang, Rong, Li, Xia, You, Lou, Jiang, Du, Meng
  et~al.}]{cao2021novel}
\bibinfo{author}{Cao, L.}, \bibinfo{author}{Yang, J.}, \bibinfo{author}{Rong,
  Z.}, \bibinfo{author}{Li, L.}, \bibinfo{author}{Xia, B.},
  \bibinfo{author}{You, C.}, \bibinfo{author}{Lou, G.}, \bibinfo{author}{Jiang,
  L.}, \bibinfo{author}{Du, C.}, \bibinfo{author}{Meng, H.}, et~al.,
  \bibinfo{year}{2021}.
\newblock \bibinfo{title}{A novel attention-guided convolutional network for
  the detection of abnormal cervical cells in cervical cancer screening}.
\newblock \bibinfo{journal}{Medical image analysis} \bibinfo{volume}{73},
  \bibinfo{pages}{102197}.
\bibitem[{Caron et~al.(2020)Caron, Misra, Mairal, Goyal, Bojanowski and
  Joulin}]{caron2020unsupervised}
\bibinfo{author}{Caron, M.}, \bibinfo{author}{Misra, I.},
  \bibinfo{author}{Mairal, J.}, \bibinfo{author}{Goyal, P.},
  \bibinfo{author}{Bojanowski, P.}, \bibinfo{author}{Joulin, A.},
  \bibinfo{year}{2020}.
\newblock \bibinfo{title}{Unsupervised learning of visual features by
  contrasting cluster assignments}.
\newblock \bibinfo{journal}{Advances in neural information processing systems}
  \bibinfo{volume}{33}, \bibinfo{pages}{9912--9924}.
\bibitem[{Caron et~al.(2021)Caron, Touvron, Misra, J{\'e}gou, Mairal,
  Bojanowski and Joulin}]{caron2021emerging}
\bibinfo{author}{Caron, M.}, \bibinfo{author}{Touvron, H.},
  \bibinfo{author}{Misra, I.}, \bibinfo{author}{J{\'e}gou, H.},
  \bibinfo{author}{Mairal, J.}, \bibinfo{author}{Bojanowski, P.},
  \bibinfo{author}{Joulin, A.}, \bibinfo{year}{2021}.
\newblock \bibinfo{title}{Emerging properties in self-supervised vision
  transformers}.
\newblock \bibinfo{journal}{arXiv preprint arXiv:2104.14294} .
\bibitem[{Chen et~al.(2020)Chen, Kornblith, Norouzi and
  Hinton}]{chen2020simple}
\bibinfo{author}{Chen, T.}, \bibinfo{author}{Kornblith, S.},
  \bibinfo{author}{Norouzi, M.}, \bibinfo{author}{Hinton, G.},
  \bibinfo{year}{2020}.
\newblock \bibinfo{title}{A simple framework for contrastive learning of visual
  representations}, in: \bibinfo{booktitle}{International conference on machine
  learning}, \bibinfo{organization}{PMLR}. pp. \bibinfo{pages}{1597--1607}.
\bibitem[{Cheng et~al.(2021)Cheng, Liu, Yu, Rao, Xiao, Han, Zhu, Lv, Li, Cai
  et~al.}]{cheng2021robust}
\bibinfo{author}{Cheng, S.}, \bibinfo{author}{Liu, S.}, \bibinfo{author}{Yu,
  J.}, \bibinfo{author}{Rao, G.}, \bibinfo{author}{Xiao, Y.},
  \bibinfo{author}{Han, W.}, \bibinfo{author}{Zhu, W.}, \bibinfo{author}{Lv,
  X.}, \bibinfo{author}{Li, N.}, \bibinfo{author}{Cai, J.}, et~al.,
  \bibinfo{year}{2021}.
\newblock \bibinfo{title}{Robust whole slide image analysis for cervical cancer
  screening using deep learning}.
\newblock \bibinfo{journal}{Nature communications} \bibinfo{volume}{12},
  \bibinfo{pages}{1--10}.
\bibitem[{Dosovitskiy et~al.(2020)Dosovitskiy, Beyer, Kolesnikov, Weissenborn,
  Zhai, Unterthiner, Dehghani, Minderer, Heigold, Gelly
  et~al.}]{dosovitskiy2020image}
\bibinfo{author}{Dosovitskiy, A.}, \bibinfo{author}{Beyer, L.},
  \bibinfo{author}{Kolesnikov, A.}, \bibinfo{author}{Weissenborn, D.},
  \bibinfo{author}{Zhai, X.}, \bibinfo{author}{Unterthiner, T.},
  \bibinfo{author}{Dehghani, M.}, \bibinfo{author}{Minderer, M.},
  \bibinfo{author}{Heigold, G.}, \bibinfo{author}{Gelly, S.}, et~al.,
  \bibinfo{year}{2020}.
\newblock \bibinfo{title}{An image is worth 16x16 words: Transformers for image
  recognition at scale}.
\newblock \bibinfo{journal}{arXiv preprint arXiv:2010.11929} .
\bibitem[{Grill et~al.(2020)Grill, Strub, Altch{\'e}, Tallec, Richemond,
  Buchatskaya, Doersch, Pires, Guo, Azar et~al.}]{grill2020bootstrap}
\bibinfo{author}{Grill, J.B.}, \bibinfo{author}{Strub, F.},
  \bibinfo{author}{Altch{\'e}, F.}, \bibinfo{author}{Tallec, C.},
  \bibinfo{author}{Richemond, P.H.}, \bibinfo{author}{Buchatskaya, E.},
  \bibinfo{author}{Doersch, C.}, \bibinfo{author}{Pires, B.A.},
  \bibinfo{author}{Guo, Z.D.}, \bibinfo{author}{Azar, M.G.}, et~al.,
  \bibinfo{year}{2020}.
\newblock \bibinfo{title}{Bootstrap your own latent: A new approach to
  self-supervised learning}.
\newblock \bibinfo{journal}{arXiv preprint arXiv:2006.07733} .
\bibitem[{He et~al.(2020)He, Fan, Wu, Xie and Girshick}]{he2020momentum}
\bibinfo{author}{He, K.}, \bibinfo{author}{Fan, H.}, \bibinfo{author}{Wu, Y.},
  \bibinfo{author}{Xie, S.}, \bibinfo{author}{Girshick, R.},
  \bibinfo{year}{2020}.
\newblock \bibinfo{title}{Momentum contrast for unsupervised visual
  representation learning}, in: \bibinfo{booktitle}{Proceedings of the IEEE/CVF
  Conference on Computer Vision and Pattern Recognition}, pp.
  \bibinfo{pages}{9729--9738}.
\bibitem[{He et~al.(2016)He, Zhang, Ren and Sun}]{he2016deep}
\bibinfo{author}{He, K.}, \bibinfo{author}{Zhang, X.}, \bibinfo{author}{Ren,
  S.}, \bibinfo{author}{Sun, J.}, \bibinfo{year}{2016}.
\newblock \bibinfo{title}{Deep residual learning for image recognition}, in:
  \bibinfo{booktitle}{Proceedings of the IEEE conference on computer vision and
  pattern recognition}, pp. \bibinfo{pages}{770--778}.
\bibitem[{Hussain et~al.(2020)Hussain, Mahanta, Das, Choudhury and
  Chowdhury}]{hussain2020shape}
\bibinfo{author}{Hussain, E.}, \bibinfo{author}{Mahanta, L.B.},
  \bibinfo{author}{Das, C.R.}, \bibinfo{author}{Choudhury, M.},
  \bibinfo{author}{Chowdhury, M.}, \bibinfo{year}{2020}.
\newblock \bibinfo{title}{A shape context fully convolutional neural network
  for segmentation and classification of cervical nuclei in pap smear images}.
\newblock \bibinfo{journal}{Artificial Intelligence in Medicine}
  \bibinfo{volume}{107}, \bibinfo{pages}{101897}.
\bibitem[{Ilse et~al.(2018)Ilse, Tomczak and Welling}]{ilse2018attention}
\bibinfo{author}{Ilse, M.}, \bibinfo{author}{Tomczak, J.},
  \bibinfo{author}{Welling, M.}, \bibinfo{year}{2018}.
\newblock \bibinfo{title}{Attention-based deep multiple instance learning}, in:
  \bibinfo{booktitle}{International conference on machine learning},
  \bibinfo{organization}{PMLR}. pp. \bibinfo{pages}{2127--2136}.
\bibitem[{Jantzen et~al.(2005)Jantzen, Norup, Dounias and
  Bjerregaard}]{jantzen2005pap}
\bibinfo{author}{Jantzen, J.}, \bibinfo{author}{Norup, J.},
  \bibinfo{author}{Dounias, G.}, \bibinfo{author}{Bjerregaard, B.},
  \bibinfo{year}{2005}.
\newblock \bibinfo{title}{Pap-smear benchmark data for pattern classification}.
\newblock \bibinfo{journal}{Nature inspired Smart Information Systems (NiSIS
  2005)} , \bibinfo{pages}{1--9}.
\bibitem[{von Karsa et~al.(2015)von Karsa, Arbyn, De~Vuyst, Dillner, Dillner,
  Franceschi, Patnick, Ronco, Segnan, Suonio et~al.}]{von2015european}
\bibinfo{author}{von Karsa, L.}, \bibinfo{author}{Arbyn, M.},
  \bibinfo{author}{De~Vuyst, H.}, \bibinfo{author}{Dillner, J.},
  \bibinfo{author}{Dillner, L.}, \bibinfo{author}{Franceschi, S.},
  \bibinfo{author}{Patnick, J.}, \bibinfo{author}{Ronco, G.},
  \bibinfo{author}{Segnan, N.}, \bibinfo{author}{Suonio, E.}, et~al.,
  \bibinfo{year}{2015}.
\newblock \bibinfo{title}{European guidelines for quality assurance in cervical
  cancer screening. summary of the supplements on hpv screening and
  vaccination}.
\newblock \bibinfo{journal}{Papillomavirus Research} \bibinfo{volume}{1},
  \bibinfo{pages}{22--31}.
\bibitem[{Kholov{\'a} et~al.(2022)Kholov{\'a}, Negri, Nasioutziki, Ventura,
  Capitanio, Bongiovanni, Cross, Bourgain, Edvardsson, Granados
  et~al.}]{kholova2022inter}
\bibinfo{author}{Kholov{\'a}, I.}, \bibinfo{author}{Negri, G.},
  \bibinfo{author}{Nasioutziki, M.}, \bibinfo{author}{Ventura, L.},
  \bibinfo{author}{Capitanio, A.}, \bibinfo{author}{Bongiovanni, M.},
  \bibinfo{author}{Cross, P.A.}, \bibinfo{author}{Bourgain, C.},
  \bibinfo{author}{Edvardsson, H.}, \bibinfo{author}{Granados, R.}, et~al.,
  \bibinfo{year}{2022}.
\newblock \bibinfo{title}{Inter-and intraobserver agreement in whole-slide
  digital thinprep samples of low-grade squamous lesions of the cervix uteri
  with known high-risk hpv status: A multicentric international study}.
\newblock \bibinfo{journal}{Cancer Cytopathology} \bibinfo{volume}{130},
  \bibinfo{pages}{939--948}.
\bibitem[{Levy et~al.(2020)Levy, De~Preux, Kenfack, Sormani, Catarino, Tincho,
  Frund, Fouogue, Vassilakos and Petignat}]{levy2020implementing}
\bibinfo{author}{Levy, J.}, \bibinfo{author}{De~Preux, M.},
  \bibinfo{author}{Kenfack, B.}, \bibinfo{author}{Sormani, J.},
  \bibinfo{author}{Catarino, R.}, \bibinfo{author}{Tincho, E.F.},
  \bibinfo{author}{Frund, C.}, \bibinfo{author}{Fouogue, J.T.},
  \bibinfo{author}{Vassilakos, P.}, \bibinfo{author}{Petignat, P.},
  \bibinfo{year}{2020}.
\newblock \bibinfo{title}{Implementing the 3t-approach for cervical cancer
  screening in cameroon: Preliminary results on program performance}.
\newblock \bibinfo{journal}{Cancer medicine} \bibinfo{volume}{9},
  \bibinfo{pages}{7293--7300}.
\bibitem[{Li et~al.(2023)Li, Liu, Liu and Liang}]{li2023novel}
\bibinfo{author}{Li, G.}, \bibinfo{author}{Liu, Q.}, \bibinfo{author}{Liu, H.},
  \bibinfo{author}{Liang, Y.}, \bibinfo{year}{2023}.
\newblock \bibinfo{title}{A novel transformer-based pipeline for lung
  cytopathological whole slide image classification}, in:
  \bibinfo{booktitle}{ICASSP 2023-2023 IEEE International Conference on
  Acoustics, Speech and Signal Processing (ICASSP)},
  \bibinfo{organization}{IEEE}. pp. \bibinfo{pages}{1--5}.
\bibitem[{Li et~al.(2019)Li, Li et~al.}]{li2019detection}
\bibinfo{author}{Li, X.}, \bibinfo{author}{Li, Q.}, et~al.,
  \bibinfo{year}{2019}.
\newblock \bibinfo{title}{Detection and classification of cervical exfoliated
  cells based on faster r-cnn}, in: \bibinfo{booktitle}{2019 IEEE 11th
  international conference on advanced infocomm technology (ICAIT)},
  \bibinfo{organization}{IEEE}. pp. \bibinfo{pages}{52--57}.
\bibitem[{Liang et~al.(2021)Liang, Pan, Sun, Liu and Du}]{liang2021global}
\bibinfo{author}{Liang, Y.}, \bibinfo{author}{Pan, C.}, \bibinfo{author}{Sun,
  W.}, \bibinfo{author}{Liu, Q.}, \bibinfo{author}{Du, Y.},
  \bibinfo{year}{2021}.
\newblock \bibinfo{title}{Global context-aware cervical cell detection with
  soft scale anchor matching}.
\newblock \bibinfo{journal}{Computer Methods and Programs in Biomedicine}
  \bibinfo{volume}{204}, \bibinfo{pages}{106061}.
\bibitem[{Lin et~al.(2019)Lin, Hu, Chen, Yao and Zhang}]{lin2019fine}
\bibinfo{author}{Lin, H.}, \bibinfo{author}{Hu, Y.}, \bibinfo{author}{Chen,
  S.}, \bibinfo{author}{Yao, J.}, \bibinfo{author}{Zhang, L.},
  \bibinfo{year}{2019}.
\newblock \bibinfo{title}{Fine-grained classification of cervical cells using
  morphological and appearance based convolutional neural networks}.
\newblock \bibinfo{journal}{IEEE Access} \bibinfo{volume}{7},
  \bibinfo{pages}{71541--71549}.
\bibitem[{Lu et~al.(2021)Lu, Williamson, Chen, Chen, Barbieri and
  Mahmood}]{lu2021data}
\bibinfo{author}{Lu, M.Y.}, \bibinfo{author}{Williamson, D.F.},
  \bibinfo{author}{Chen, T.Y.}, \bibinfo{author}{Chen, R.J.},
  \bibinfo{author}{Barbieri, M.}, \bibinfo{author}{Mahmood, F.},
  \bibinfo{year}{2021}.
\newblock \bibinfo{title}{Data-efficient and weakly supervised computational
  pathology on whole-slide images}.
\newblock \bibinfo{journal}{Nature biomedical engineering} \bibinfo{volume}{5},
  \bibinfo{pages}{555--570}.
\bibitem[{Van~der Maaten and Hinton(2008)}]{van2008visualizing}
\bibinfo{author}{Van~der Maaten, L.}, \bibinfo{author}{Hinton, G.},
  \bibinfo{year}{2008}.
\newblock \bibinfo{title}{Visualizing data using t-sne.}
\newblock \bibinfo{journal}{Journal of machine learning research}
  \bibinfo{volume}{9}.
\bibitem[{Organization et~al.(2021)}]{world2021guideline}
\bibinfo{author}{Organization, W.H.}, et~al., \bibinfo{year}{2021}.
\newblock \bibinfo{title}{WHO guideline for screening and treatment of cervical
  pre-cancer lesions for cervical cancer prevention}.
\newblock \bibinfo{publisher}{World Health Organization}.
\bibitem[{Paszke et~al.(2019)Paszke, Gross, Massa, Lerer, Bradbury, Chanan,
  Killeen, Lin, Gimelshein, Antiga et~al.}]{paszke2019pytorch}
\bibinfo{author}{Paszke, A.}, \bibinfo{author}{Gross, S.},
  \bibinfo{author}{Massa, F.}, \bibinfo{author}{Lerer, A.},
  \bibinfo{author}{Bradbury, J.}, \bibinfo{author}{Chanan, G.},
  \bibinfo{author}{Killeen, T.}, \bibinfo{author}{Lin, Z.},
  \bibinfo{author}{Gimelshein, N.}, \bibinfo{author}{Antiga, L.}, et~al.,
  \bibinfo{year}{2019}.
\newblock \bibinfo{title}{Pytorch: An imperative style, high-performance deep
  learning library}.
\newblock \bibinfo{journal}{Advances in neural information processing systems}
  \bibinfo{volume}{32}, \bibinfo{pages}{8026--8037}.
\bibitem[{P{\'e}rez et~al.(2003)P{\'e}rez, Gangnet and
  Blake}]{perez2003poisson}
\bibinfo{author}{P{\'e}rez, P.}, \bibinfo{author}{Gangnet, M.},
  \bibinfo{author}{Blake, A.}, \bibinfo{year}{2003}.
\newblock \bibinfo{title}{Poisson image editing}, in: \bibinfo{booktitle}{ACM
  SIGGRAPH 2003 Papers}, pp. \bibinfo{pages}{313--318}.
\bibitem[{Plissiti et~al.(2018)Plissiti, Dimitrakopoulos, Sfikas, Nikou,
  Krikoni and Charchanti}]{8451588}
\bibinfo{author}{Plissiti, M.E.}, \bibinfo{author}{Dimitrakopoulos, P.},
  \bibinfo{author}{Sfikas, G.}, \bibinfo{author}{Nikou, C.},
  \bibinfo{author}{Krikoni, O.}, \bibinfo{author}{Charchanti, A.},
  \bibinfo{year}{2018}.
\newblock \bibinfo{title}{Sipakmed: A new dataset for feature and image based
  classification of normal and pathological cervical cells in pap smear
  images}, in: \bibinfo{booktitle}{2018 25th IEEE International Conference on
  Image Processing (ICIP)}, pp. \bibinfo{pages}{3144--3148}.
\newblock \DOIprefix\doi{10.1109/ICIP.2018.8451588}.
\bibitem[{Redmon and Farhadi(2018)}]{redmon2018yolov3}
\bibinfo{author}{Redmon, J.}, \bibinfo{author}{Farhadi, A.},
  \bibinfo{year}{2018}.
\newblock \bibinfo{title}{Yolov3: An incremental improvement}.
\newblock \bibinfo{journal}{arXiv preprint arXiv:1804.02767} .
\bibitem[{Saidu et~al.(2021)Saidu, Kuhn, Tergas, Boa, Moodley, Svanholm-Barrie,
  Persing, Campbell, Tsai, Wright et~al.}]{saidu2021performance}
\bibinfo{author}{Saidu, R.}, \bibinfo{author}{Kuhn, L.},
  \bibinfo{author}{Tergas, A.}, \bibinfo{author}{Boa, R.},
  \bibinfo{author}{Moodley, J.}, \bibinfo{author}{Svanholm-Barrie, C.},
  \bibinfo{author}{Persing, D.}, \bibinfo{author}{Campbell, S.},
  \bibinfo{author}{Tsai, W.Y.}, \bibinfo{author}{Wright, T.C.}, et~al.,
  \bibinfo{year}{2021}.
\newblock \bibinfo{title}{Performance of xpert hpv on self-collected vaginal
  samples for cervical cancer screening among women in south africa}.
\newblock \bibinfo{journal}{Journal of lower genital tract disease}
  \bibinfo{volume}{25}, \bibinfo{pages}{15}.
\bibitem[{Shao et~al.(2021)Shao, Bian, Chen, Wang, Zhang, Ji and
  Zhang}]{shao2021transmil}
\bibinfo{author}{Shao, Z.}, \bibinfo{author}{Bian, H.}, \bibinfo{author}{Chen,
  Y.}, \bibinfo{author}{Wang, Y.}, \bibinfo{author}{Zhang, J.},
  \bibinfo{author}{Ji, X.}, \bibinfo{author}{Zhang, Y.}, \bibinfo{year}{2021}.
\newblock \bibinfo{title}{Transmil: Transformer based correlated multiple
  instance learning for whole slide image classication}.
\newblock \bibinfo{journal}{arXiv preprint arXiv:2106.00908} .
\bibitem[{Stegm{\"u}ller et~al.(2023)Stegm{\"u}ller, Bozorgtabar, Spahr and
  Thiran}]{stegmuller2023scorenet}
\bibinfo{author}{Stegm{\"u}ller, T.}, \bibinfo{author}{Bozorgtabar, B.},
  \bibinfo{author}{Spahr, A.}, \bibinfo{author}{Thiran, J.P.},
  \bibinfo{year}{2023}.
\newblock \bibinfo{title}{Scorenet: Learning non-uniform attention and
  augmentation for transformer-based histopathological image classification},
  in: \bibinfo{booktitle}{Proceedings of the IEEE/CVF Winter Conference on
  Applications of Computer Vision}, pp. \bibinfo{pages}{6170--6179}.
\bibitem[{Sung et~al.(2021)Sung, Ferlay, Siegel, Laversanne, Soerjomataram,
  Jemal and Bray}]{sung2021global}
\bibinfo{author}{Sung, H.}, \bibinfo{author}{Ferlay, J.},
  \bibinfo{author}{Siegel, R.L.}, \bibinfo{author}{Laversanne, M.},
  \bibinfo{author}{Soerjomataram, I.}, \bibinfo{author}{Jemal, A.},
  \bibinfo{author}{Bray, F.}, \bibinfo{year}{2021}.
\newblock \bibinfo{title}{Global cancer statistics 2020: Globocan estimates of
  incidence and mortality worldwide for 36 cancers in 185 countries}.
\newblock \bibinfo{journal}{CA: a cancer journal for clinicians}
  \bibinfo{volume}{71}, \bibinfo{pages}{209--249}.
\bibitem[{Touvron et~al.(2021)Touvron, Cord, Douze, Massa, Sablayrolles and
  Jegou}]{pmlr-v139-touvron21a}
\bibinfo{author}{Touvron, H.}, \bibinfo{author}{Cord, M.},
  \bibinfo{author}{Douze, M.}, \bibinfo{author}{Massa, F.},
  \bibinfo{author}{Sablayrolles, A.}, \bibinfo{author}{Jegou, H.},
  \bibinfo{year}{2021}.
\newblock \bibinfo{title}{Training data-efficient image transformers \&
  distillation through attention}, in: \bibinfo{editor}{Meila, M.},
  \bibinfo{editor}{Zhang, T.} (Eds.), \bibinfo{booktitle}{Proceedings of the
  38th International Conference on Machine Learning},
  \bibinfo{publisher}{PMLR}. pp. \bibinfo{pages}{10347--10357}.
\newblock \URLprefix \url{https://proceedings.mlr.press/v139/touvron21a.html}.
\bibitem[{Vassilakos et~al.(2023)Vassilakos, Clarke, Murtas, Stegm{\"u}ller,
  Wisniak, Akhoundova, Sando, Orock, Sormani, Thiran
  et~al.}]{vassilakos2023telecytologic}
\bibinfo{author}{Vassilakos, P.}, \bibinfo{author}{Clarke, H.},
  \bibinfo{author}{Murtas, M.}, \bibinfo{author}{Stegm{\"u}ller, T.},
  \bibinfo{author}{Wisniak, A.}, \bibinfo{author}{Akhoundova, F.},
  \bibinfo{author}{Sando, Z.}, \bibinfo{author}{Orock, G.E.},
  \bibinfo{author}{Sormani, J.}, \bibinfo{author}{Thiran, J.P.}, et~al.,
  \bibinfo{year}{2023}.
\newblock \bibinfo{title}{Telecytologic diagnosis of cervical smears for triage
  of self-sampled human papillomavirus--positive women in a resource-limited
  setting: concept development before implementation}.
\newblock \bibinfo{journal}{Journal of the American Society of Cytopathology} .
\bibitem[{Wei et~al.(2021)Wei, Cheng, Liu and Zeng}]{wei2021efficient}
\bibinfo{author}{Wei, Z.}, \bibinfo{author}{Cheng, S.}, \bibinfo{author}{Liu,
  X.}, \bibinfo{author}{Zeng, S.}, \bibinfo{year}{2021}.
\newblock \bibinfo{title}{An efficient cervical whole slide image analysis
  framework based on multi-scale semantic and spatial deep features}.
\newblock \bibinfo{journal}{arXiv preprint arXiv:2106.15113} .
\bibitem[{Zhou et~al.(2021)Zhou, Wei, Wang, Shen, Xie, Yuille and
  Kong}]{zhou2021ibot}
\bibinfo{author}{Zhou, J.}, \bibinfo{author}{Wei, C.}, \bibinfo{author}{Wang,
  H.}, \bibinfo{author}{Shen, W.}, \bibinfo{author}{Xie, C.},
  \bibinfo{author}{Yuille, A.}, \bibinfo{author}{Kong, T.},
  \bibinfo{year}{2021}.
\newblock \bibinfo{title}{ibot: Image bert pre-training with online tokenizer}.
\newblock \bibinfo{journal}{arXiv preprint arXiv:2111.07832} .

\end{thebibliography}

\end{document}